\documentclass[useAMS,usenatbib]{mnras}
\usepackage{aasmacros,placeins,float,times}
\usepackage{amsmath}
\usepackage{graphicx,url}
\usepackage{amssymb}

\usepackage[usenames,dvipsnames]{color}
\usepackage[normalem]{ulem}
\title[Halo properties in the COCO WDM simulation]{The COpernicus COmplexio: Statistical Properties of Warm Dark Matter Haloes}
\author[S. Bose et al.]{Sownak Bose$^{1}$\thanks{E-mail:
   sownak.bose@durham.ac.uk},  Wojciech A. Hellwing$^{1,2}$,
Carlos  S. Frenk$^{1}$, Adrian Jenkins$^{1}$, \newauthor Mark R. Lovell$^{3,4}$, John C. Helly$^{1}$ and Baojiu Li$^{1}$ \\
 $^{1}$Institute for Computational Cosmology, Durham University, South Road, Durham, UK, DH1 3LE\\
$^{2}$ICM, University of Warsaw, ul. Pawi{\'n}skiego 5a, 02-106 Warsaw, Poland \\
$^{3}$GRAPPA Institute, Universiteit van Amsterdam, Science Park 904, 1098 XH Amsterdam, The Netherlands \\
$^{4}$Instituut-Lorentz for Theoretical Physics, Niels Bohrweg 2, NL-2333 CA Leiden, The Netherlands \\}

\newcommand{\Msun}{h^{-1}\,M_\odot}

\def\gsim{ \lower .75ex \hbox{$\sim$} \llap{\raise .27ex \hbox{$>$}} }
\def\lsim{ \lower .75ex \hbox{$\sim$} \llap{\raise .27ex \hbox{$<$}} }

\def\jcap{JCAP}
\newcommand{\bq}{\begin{eqnarray}}
\newcommand{\eq}{\end{eqnarray}}

\def\eg{{\frenchspacing\it e.g.}}

\def\hmpc{h^{-1}\,{\rm Mpc}}

\begin{document}

\date{Accepted *** Received ***; in original
    form ***} 
    
    \pagerange{\pageref{firstpage}--\pageref{lastpage}} \pubyear{2015}

  \maketitle

  \label{firstpage}
  
\begin{abstract}
  The recent detection of a 3.5 keV X-ray line from the centres of
  galaxies and clusters by \cite{2014ApJ...789...13B} and
  \cite{2014PhRvL.113y1301B} has been interpreted as emission from the
  decay of 7~keV sterile neutrinos which could make up the (warm) dark
  matter (WDM).  As part of the {\em COpernicus COmplexio}
  (\textsc{coco}) programme, we investigate the properties of dark
  matter haloes formed in a high-resolution cosmological $N$-body
  simulation from initial conditions similar to those expected in a
  universe in which the dark matter consists of 7~keV sterile
  neutrinos. This simulation and its cold dark matter (CDM)
  counterpart have $\sim13.4$bn particles, each of mass $\sim 10^5\,
  \Msun$, providing detailed information about halo structure and
  evolution down to dwarf galaxy mass scales.  Non-linear structure
  formation on small scales ($M_{200}\, \lsim\, 2 \times 10^9~h^{-1}\,M_\odot$)
  begins slightly later in \textsc{coco-warm} than in
  \textsc{coco-cold}.  The halo mass function at the present day in
  the WDM model begins to drop below its CDM counterpart at a mass
  $\sim 2 \times 10^{9}~h^{-1}\,M_\odot$ and declines very rapidly
  towards lower masses so that there are five times fewer haloes of
  mass $M_{200}= 10^{8}~h^{-1}\,M_\odot$ in \textsc{coco-warm} than in
  \textsc{coco-cold}.  Halo concentrations on dwarf galaxy scales are
  correspondingly smaller in \textsc{coco-warm}, and we provide a
  simple functional form that describes its evolution with
  redshift. The shapes of haloes are similar in the two cases, but the
  smallest haloes in \textsc{coco-warm} rotate slightly more slowly
  than their CDM counterparts.
\end{abstract}
 
\section{Introduction}
\label{intro}
 
The identity of dark matter, the dominant matter component of the Universe, has long been a subject of great interest in cosmology. In the last three decades, the model of non-relativistic dark matter consisting of heavy weakly-interacting particles with negligible thermal velocities at early times, the Cold Dark Matter (CDM) model, has become the cornerstone of the standard cosmological paradigm. The standard model with dark energy in the form of a cosmological constant, $\Lambda$ ($\Lambda$CDM, henceforth just CDM) has been very successful in predicting and matching observational data on a wide range of scales, from the temperature fluctuations in the Cosmic Microwave Background (\citealt{2014A&A...571A...1P}) to the statistics of galaxy clustering (\citealt{2dfgrs,Zehavi2002,Hawkins2003,sdss,bao_2df,bao_sdss}; for a comprehensive review on the subject, see \citealt{2012AnP...524..507F}).

With the advent of the LHC it was hoped that one of the best-motivated CDM candidates, the lightest supersymmetric particle (the neutralino) would be found. The lack of evidence for supersymmetry at the LHC and the absence of a convincing direct or indirect signal for CDM (but see \citealt{2011PhLB..697..412H}) has encouraged the exploration of viable alternatives. One of the most promising alternatives is the sterile neutrino (\citealt{1994PhRvL..72...17D,2005PhLB..620...17A}), which behaves as warm dark matter (WDM) due to the particles' non-negligible thermal velocities at early times. Being collisionless, this leads to free streaming and the damping of perturbations in the density field, creating a cutoff in the matter power spectrum on the scale of dwarf galaxies.   

A simple extension of the Standard Model of particle physics, called the neutrino Minimal Standard Model ($\nu$MSM, \citealt{2009ARNPS..59..191B}), consists of three right-handed sterile neutrinos in which, for a specific choice of parameters, one of the sterile neutrinos behaves as a dark matter particle and the model explains neutrino flavour oscillations. Each one of this triplet of particles has its mass below the electroweak scale; one in the keV scale (denoted by $M_1$), and two in the GeV scale (denoted by $M_2$ and $M_3$). The former behaves as a relativistic particle at the time of neutrino decoupling and acts as WDM, and is then redshifted to non-relativistic energies during the radiation-dominated era. Unlike a thermal relic, the cutoff in the power spectrum introduced by a sterile neutrino of a fixed mass depends on a second parameter, the lepton asymmetry. As we explain later in the following section, it is possible to approximate the sterile neutrino power spectrum with a WDM thermal relic equivalent, particularly for very low and very high values of the lepton asymmetry.

The unidentified 3.53 keV X-ray line originally detected in the spectrum of a stack of galaxy clusters (\citealt{2014ApJ...789...13B}) and in the spectra of M31 and the Perseus cluster (\citealt{2014PhRvL.113y1301B}) could be a decay signal of sterile neutrino dark matter, with a particle mass of 7 keV. More recently, \cite{2014arXiv1408.2503B} have also identified a similar line in the centre of the Milky Way. There are, however, several groups that have questioned the plausibility of this detection. For example, \cite{2014arXiv1405.7943R} failed to find a signal in Chandra observations of the Milky Way. Of course, the Galactic centre is heavily contaminated by X-rays, which introduces uncertainties, a point made by \cite{2014arXiv1408.2503B}. 

Systematic effects can result from the atomic data used in modelling the plasma, as argued by \cite{2014arXiv1408.1699J}, who found no excess when re-analysing the \cite{2014PhRvL.113y1301B} data and claimed that any signal at 3.5 keV could be explained by known Potassium (K XVIII) and Chlorine (Cl XVII) lines. \cite{2014arXiv1409.4143B} put this latter result down to the use of ``incorrect atomic data and inconsistent spectroscopic modelling'' by \cite{2014arXiv1408.1699J}. A further non-detection was then reported in the stacked spectra of galaxies from Chandra and XMM-Newton (\citealt{Anderson}), while most recently, \cite{2014PhRvD..90j3506M} analysed the spectra of stacked dwarf galaxies from XMM-Newton and claimed to rule out the Andromeda signal detected by \cite{2014ApJ...789...13B} at the $4.6\sigma$ confidence level. This has spurred other groups (see, for example, \citealt{2014JCAP...11..033C}) to associate the 3.53 keV signals to the conversion of a sterile neutrino into an axion, and its subsequent decay into photons. Such a mechanism requires a magnetic field, the presence and strength of which can vary from galaxy to galaxy, a scenario that could explain why this line is only seen in some objects.

Clearly, whether or not the 3.53 keV line really does exist remains an open question. It is, therefore, important to investigate the predictions for the formation of cosmic structures in a model in which the dark matter consists of particles that could decay producing such a line. Constraints on such models can be set from the observed clustering of the Lyman-$\alpha$ forest at high redshift whose small-scale structure would be erased if the dark matter were warm. On these grounds, \cite{2013PhRvD..88d3502V} recently set a (current) lower limit of 3.3 keV for the mass of a dominant thermal warm dark matter particle.

Coincidentally, the power spectrum of a $3.3$ keV thermal warm dark
matter particle is well approximated by that of a $7$ keV sterile
neutrino for a lepton asymmetry of $L_6=8.66$. This corresponds to the
smallest allowed value of the power spectrum cutoff length (i.e. to
the ``coldest'' power spectrum possible) for a sterile neutrino of
mass $7$~keV. This is the model that we will explore in this
work. Ruling out this model from astronomical data on small scales
would rule out the entire family of $7$~keV sterile neutrino
candidates. To investigate the model we use high resolution $N$-body
simulations whose results we compare with those of CDM simulations
with the same phases in the initial conditions. We are interested
exclusively in characterising the properties of dark matter haloes of
mass in the region of the power spectrum cutoff and, in this study, we
ignore the effects of baryons. Such effects must be taken into account
when comparing model predictions with observations. In the case of
CDM, relevant baryon effects on the small scales of interest here have
recently been quantified by
\citealt{2013MNRAS.431.1366S,2014arXiv1406.6362S,2015MNRAS.448.2941S,2015MNRAS.451.1247S})

The layout of this paper is as follows. In Section~\ref{SimOverall},
we introduce the simulations used in this work, the modelling of the
WDM component, and describe how we tackle the issue of spurious halo
formation in our simulations. In Section~\ref{Results} we present our
main results from the comparison of WDM and CDM from our simulations,
in terms of both the large-scale distribution of matter, and the
internal structure of haloes. Finally,
in Section~\ref{conclusion}, we summarise our findings and look into
some future work that will be carried out with the same set of
simulations.
\section{The Simulations}
\label{SimOverall}

In this section, we provide an overview of the initial conditions and modelling of the WDM component in our simulations.

\subsection{The simulation set-up}
\label{SimParams}

The $N$-body simulations presented in this paper are part of the {\it
  COpernicus COmplexio} (\textsc{coco}) simulation programme
\citep{Hellwing_15} being carried out by the {\it Virgo Consortium}.
This is a set of cosmological ``zoom-in'' simulations
(\citealt{1993ApJ...412..455K,Frenk_1996}), as was done in the
\textsc{gimic} simulations (\citealt{2009MNRAS.399.1773C}). The parent
simulation, called the {\it COpernicus complexio LOw Resolution} (or
\textsc{color}) simulation, followed the evolution of $4.25$ billion
particles in a periodic box of size $70.4\, h^{-1}~$Mpc. We extracted a
roughly spherical region of radius $\sim 18\hmpc$, and centred on the
location $(42.2, 51.2, 8.8)~h^{-1}\,\mathrm{Mpc}$ in the
\textsc{color} volume. Both \textsc{color} and \textsc{coco} assume
cosmological parameters derived from the seven-year {\it Wilkinson
  Microwave Anisotropy Probe} (WMAP 7) data
(\citealt{2011ApJS..192...18K}), with the parameters: $\Omega_m =
0.272$, $\Omega_\Lambda = 0.728$, $h = 0.704$, $n_s = 0.967$ and
$\sigma_8 = 0.81$. Here, $\Omega_{\{m,\Lambda\}}$ represents the
present-day fractional contribution of matter and the cosmological
constant respectively, in units of the critical density $\rho_c =
3H_0^2 / 8\upi G$, $h = H_0 / 100\,\mathrm{km/s/Mpc}$ is the
dimensionless Hubble parameter, $n_s$ is the spectral index of the
primordial power spectrum, and $\sigma_8$ is the linear {\it rms}
density fluctuation in a sphere of radius $8\,h^{-1}$ Mpc at $z = 0$.

Dark matter particles with three different masses are used in regions
simulated at different resolutions within the parent simulation
volume. Initially, the high-resolution region has a shape similar to
an amoeba which approximates a sphere of radius $\sim 17.4\hmpc$ at
the present time. It contains $12.9$ billion particles of mass $1.135
\times 10^5\, \Msun$. The volume surrounding this region contains the
medium- ($3.07 \times 10^{6}\, \Msun$) and low-resolution ($1.96
\times 10^8\Msun$) particles. We have taken care to minimise
contamination of the high-resolution region by lower mass particles
and all the haloes discussed in this study are entirely made up of the
high-resolution particles. The gravitational softening was kept fixed
at $\epsilon \sim 230\, h^{-1}$ pc for the high-resolution particles,
increasing by a factor of 10 each time for the medium- and
low-resolution particles.

The simulation ran from $z = 127$ to $z = 0$ using the
\textsc{gadget3} code, which is an updated version of the publicly
available \textsc{gadget2} code \citep{GADGET,GADGET2}. Phase
information for the creation of the initial conditions for both
\textsc{coco-warm} and \textsc{coco-cold} was obtained from the public
Gaussian white noise field \textsc{panphasia}
(\citealt{2013MNRAS.434.2094J}), and perturbations thereafter were
calculated using the second-order Lagrangian Perturbation Theory
(2LPT) algorithm presented in \cite{2010MNRAS.403.1859J}. The details
of the simulation, along with the \textsc{panphasia} phase descriptor,
are summarised in Table~\ref{simdata}.

\begin{table*}
\centering
\vspace{10pt}
\begin{tabular}{ccc}
\hline \hline \\
Parameter & \textsc{color}  (Parent volume) & \textsc{coco} (This paper)\\ \\
\hline \hline \\
Box Size & 70.4 $h^{-1}$ Mpc & $-$\\
$m_{\mathrm{WDM}}$ & 3.3 keV & $-$ \\
$N_p$ & $4,251,528,000$ & $13, 384, 245, 248$\\
$V_{\mathrm{hr}}$ & $70.4^3~h^{-3}$Mpc$^3$  & $\sim2.2\times 10^4 ~h^{-3}$Mpc$^3$  \\
$m_{p, \mathrm{hr}}$  & $6.196 \times 10^6~h^{-1}\,M_\odot$ & $1.135 \times 10^5~h^{-1}\,M_\odot$\\
$N_{p, \mathrm{hr}}$  & $4,251,528,000$ & $12,876,807,168$ \\
$\epsilon_{\mathrm{hr}}$ & 1 $h^{-1}$ kpc & 230 $h^{-1}$ pc \\
$h$ & 0.704 & $-$ \\
$\Omega_m$ & 0.272 & $-$ \\
$\Omega_{\Lambda}$ & 0.728 & $-$ \\
$\sigma_8$ & 0.81 & $-$\\
Phase descriptor & [Panph1,L16,(31250,23438,39063),S12,CH582187950,COLOR] & $-$ \\
\\ \hline \hline
\end{tabular}
\caption{Cosmological parameters used in the \textsc{coco} simulations, and its parent volume, \textsc{color}. Here, $m_{\mathrm{WDM}}$ is the mass of the thermal relic warm dark matter particle, $N_p$ is the total number of particles (of all types) used in the simulation,$V_{\mathrm{hr}}$ is the approximate volume of the high-resolution region at $z=0$, $m_{p, \mathrm{hr}}$ is the mass of an individual high-resolution dark matter particle, $N_{p, \mathrm{hr}}$ is the total number of particles of this species, whereas $\epsilon_{\mathrm{hr}}$ is the softening length applied to them. The cosmological parameters $h, \Omega_m, \Omega_\Lambda$ and $\sigma_8$ are as described in the text. The phases for the parent \textsc{color} simulation can be generated using the \textsc{panphasia} phase descriptor provided in the last row. The blank fields in the \textsc{coco} column mean that the parameter assumes the same value as in the parent simulation, \textsc{color}.}
\label{simdata}
\end{table*}

The distinctive feature of WDM particles are non-negligible thermal
velocities at early times, which result in free streaming that washes
out perturbations in the matter distribution below the free streaming
scale~(\citealt{1983ApJ...274..443B,2012MNRAS.424..684S,2013MNRAS.428.1774B}).
As a result, we expect the abundance, distribution and internal
structure of WDM haloes to be different from those of CDM
haloes. Indeed, thermal velocities introduce a limit to the
fine-grained phase space density in dark matter haloes, creating cores
in the density profile (\citealt{ 2012MNRAS.424.1105M,2013MNRAS.430.2346S}). 
However, as shown in these papers, the cores produced by realistic thermal relics are only a few parsecs in size,
and thus not astrophysically relevant. In our simulations we can
neglect these thermal velocities, which at $z = 0$ are of the order of
a few tens of metres per second \citep{2012MNRAS.420.2318L} so, over the course of the
simulation, which starts at $z = 127$, the particles would travel only
a few kiloparsecs, comparable to the mean interparticle spacing of the
high-resolution particles.

The WDM power spectrum of density fluctuations is often modelled by
the transfer function, $T(k)$, relative to the CDM case:
\bq
P_{\mathrm{WDM}} (k) = T^2(k) P_{\mathrm{CDM}} (k)\;.
\eq
We approximate $T(k)$ using the fitting formula provided by \cite{2001ApJ...556...93B}:
\bq \label{transfer}
T(k) = \left(1 + \left(\alpha k \right)^{2 \nu} \right)^{-5 / \nu}\;,
\eq
where $\alpha$ and $\nu$ are constants.  As computed by \cite{2005PhRvD..71f3534V}, for  $k < 5~h^{-1}\,\mathrm{Mpc}$, the value $\nu = 1.12$ provides the best-fitting transfer function. The value of $\alpha$ is dependent on the mass of the WDM particle (\citealt{2005PhRvD..71f3534V}):
\bq \label{alphaMass}
\alpha = 0.049 \left[ \frac{m_{\mathrm{{\tiny WDM}}}}{\mathrm{keV}} \right]^{-1.11} \left[ \frac{\Omega_{\mathrm{{\tiny WDM}}}}{0.25} \right]^{0.11}\left[\frac{h}{0.7}\right]~h^{-1}~\mathrm{Mpc},
\eq
and determines the scale of the cutoff due to free streaming in the
WDM power spectrum relative to CDM. It should be noted that this
transfer function is a fit to the full thermal relic power spectrum,
obtained by solving the Boltzmann equation (Lovell et al. in
prep.).

As we can see in Eq.~\ref{alphaMass}, the ``warmer" the dark matter
particle (i.e., the lower its rest mass is), the larger the scale at
which the cutoff in the power spectrum occurs. 
One way to define the characteristic scale in the
power spectrum is through 
the ``half-mode'' wavenumber, $k_{\mathrm{hm}}$, where the transfer
function in Eq.~\ref{transfer} drops by a factor of two:
\bq
k_{\mathrm{hm}} = \frac{1}{\alpha} \left( 2^{\nu/5} -1 \right)^{1/2\nu}\;.
\eq
The  associated ``half-mode mass'', $M_{\mathrm{hm}}$, is the mean density enclosed within this half-mode:
\bq
M_{\mathrm{hm}} = \frac{4\upi}{3} \bar{\rho} \left( \frac{\lambda_{\mathrm{hm}}}{2} \right)^{3}\;.
\eq
 For the 3.3 keV model, this occurs at around $M_{\mathrm{hm}}
\sim~2~\times~10^8~h^{-1}\,M_\odot$
(\citealt{2008ApJ...673..203C,2013MNRAS.434.3337A,2013PhRvD..88d3502V}).
We will show later that
differences in the formation time of haloes in WDM and CDM begin to
appear below $\sim 2 \times 10^9~h^{-1}\,M_\odot$, approximately an order of magnitude above the 
half-mode mass scale.

The form of the power spectrum for sterile neutrino dark matter is
determined by two parameters -- the mass of the particle, $m_{\nu_s}$,
and the lepton asymmetry, $L_6$:
\bq
L_6 \equiv 10^6 \left( \frac{n_{\nu_e} - n_{\bar{\nu}_e}}{s} \right), 
\eq
where $n_{\nu_e}$ is the number density in electron neutrinos,
$n_{\bar{\nu}_e}$ the number density in electron anti-neutrinos, and
$s$ is the entropy density of the Universe~(\citealt{Laine2008}). 
The scales at which the power spectrum is suppressed for sterile neutrinos vary
non-monotonically as a function of $L_6$. If $L_6$ is very small
($\ll$ 1) the power spectrum exhibits a similar abrupt cutoff to that of a
thermal relic. As $L_6$ is increased, the cutoff becomes gentler and
$k_{\mathrm{hm}}$ shifts to larger values. At some value of $L_6$ (typically
between 8 and 25 depending on the sterile neutrino mass), $k_{\mathrm{hm}}$
reaches a maximum; for still higher $L_6$, $k_{\mathrm{hm}}$
retreats to lower $k$ and returns to its original shape and position
(\citealt{1999PhRvL..82.2832S, 2014PhRvL.112p1303A}, Lovell et al. in
prep.).

\begin{figure}
\includegraphics[width=0.5\textwidth]{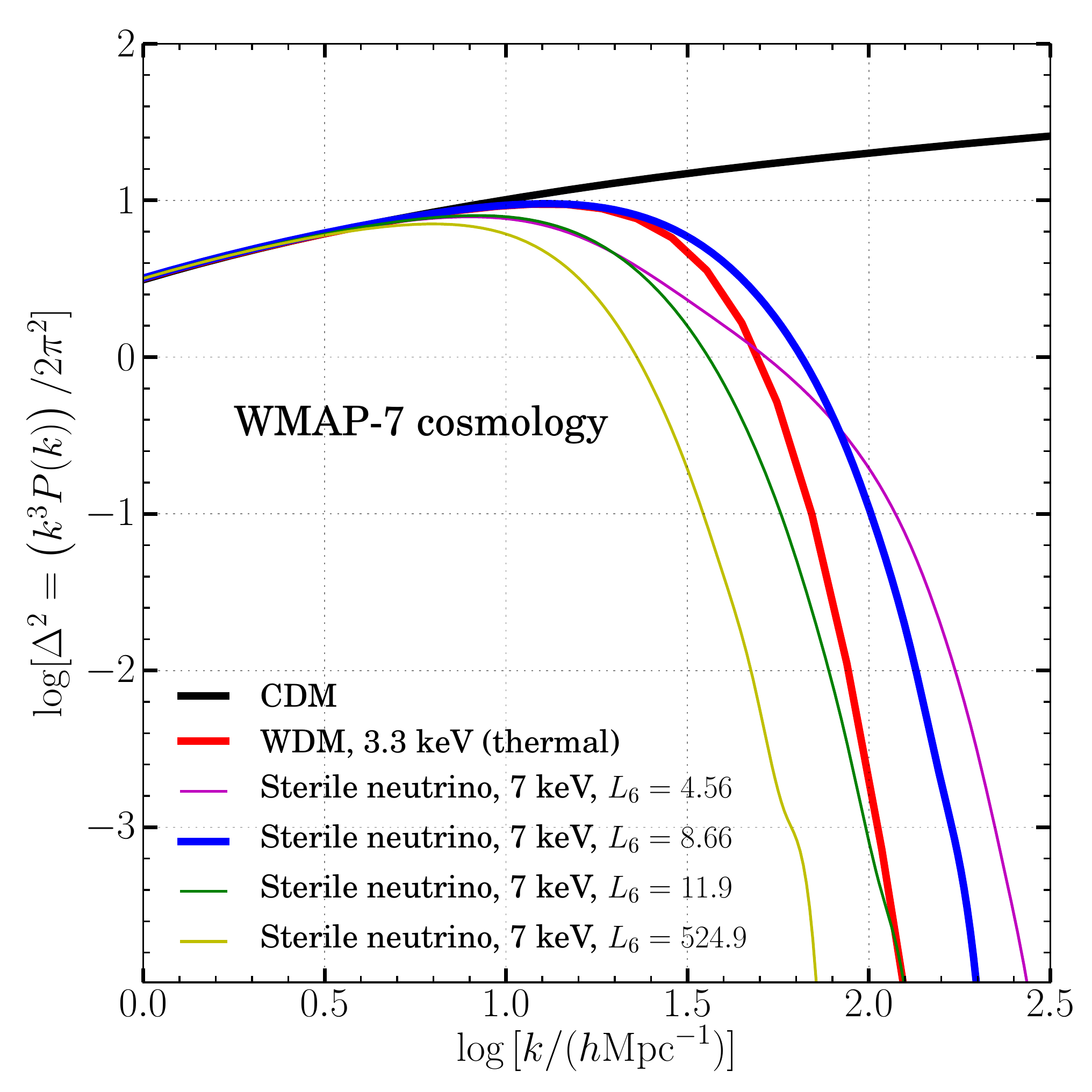}
\caption{The (dimensionless) matter power spectrum for: a thermal
  $3.3$ keV WDM (red), a sterile neutrino of mass $m_{\nu_s} = 7$ keV
  and lepton asymmetry $L_6 = 8.66$ (blue) and CDM (black). Both the
  WDM and sterile neutrino power spectra have significantly suppressed
  power at small scales, with the deviation from CDM case at almost
  identical scales: $\log(k) \gtrsim 1.0~h~\mathrm{Mpc}^{-1}$. Also
  shown as thin coloured lines are power spectra for 7~keV sterile
  neutrinos with different values of $L_6$, as indicated in the
  legend. }
\label{matterps}
\end{figure}

The power spectra used in the \textsc{coco} simulations are shown as
thick lines in Fig.~\ref{matterps}: CDM in black, 3.3 keV WDM in red
and 7 keV sterile neutrinos with $L_6 = 8.66$ in blue. All three power
spectra agree on large scales. On small scales, the two warm dark
matter models differ from CDM. $k_{\mathrm{hm}}$ for
the sterile neutrino case occurs at a very similar scale, and the cutoff has a
similar shape to that for the thermal relic case. On smaller
scales still, the sterile neutrino power spectrum has more power than
its thermal counterpart, but the differences only become significant
on scales where the amplitude is, at most, a few percent of the peak
amplitude. These differences are negligible and can be safely ignored
in our simulations. The thin lines in the figure correspond to 7 keV
sterile neutrino power spectra for different values of the lepton
asymmetry, $L_6$. The $L_6 = 8.66$ model that we have simulated
corresponds to the ``coldest'' possible 7 keV sterile neutrino.

\subsection{Halo identification and matching}
\label{matching}

Haloes were identified in our simulations using the friend-of-friend
(FOF) algorithm (\citealt{1985ApJ...292..371D}) with a linking length
of 0.2 times the mean interparticle separation, and a minimum of 20
particles. Gravitationally-bound substructures within these groups
were then identified using the {\sc subfind} algorithm
(\citealt{2001MNRAS.328..726S}), although in this paper, we will be
mostly concerned with the properties of the WDM FOF groups. We
determine the halo centre using the ``shrinking sphere" method of
\cite{2003MNRAS.338...14P}. In short, we recursively compute the
centre of mass of all particles within a shrinking sphere, until a
convergence criterion is met. In each iteration, the radius of the
sphere is reduced by $5\%$, and stopped when only 1000 particles or
$1\%$ of the particles of the initial sphere (whichever is smaller) are
left.

Comparing halo statistics between sets of simulations requires
consistent definitions for the various properties of the haloes. In
this work, we make use of two definitions of mass: $M_{\mathrm{FOF}}$,
which is the mass of all particles identified by the algorithm as
belonging to the FOF group, and $M_{200}$, which is the mass contained
within a sphere of radius $r_{200}$ (centred on the ``shrinking
sphere'' centre defined above), within which the average density is
200 times the critical density of the Universe ($\rho_c$) at the
specified redshift. Another common radius used to define a halo edge
is the virial radius, $r_{\mathrm{vir}}$, within which the density of
the halo $\bar{\rho}(<r_{\mathrm{vir}}) = \Delta \rho_c$, where $\Delta \sim 178
\Omega_m^{0.45}$ (motivated by the spherical collapse model,
\citealt{1996MNRAS.282..263E}. Note that this definition is consistent
with the virial overdensity relation
in~\citealt{Bryan1998}). Table~\ref{halonums} summarises the total
number of groups and self-bound substructures identified at $z = 0$ in
our simulations.

\begin{table}
\centering
\vspace{10pt}
\begin{tabular}{ccc}
\hline \hline \\
Simulation & $N_{\mathrm{FOF}} (z=0)$ & $N_{\mathrm{subs}} (z=0)$\\ \\
\hline \hline \\
\textsc{color-cold} & $3,961,192$ & $4,770,041$ \\
\textsc{color-warm} & $2,609,122$ & $3,082,275$ \\
\textsc{coco-cold} & $8,896,811$ & $10,502,187$ \\
\textsc{coco-warm} & $2,548,743 $ & $2,830,514$ \\
\\ \hline \hline
\end{tabular}
\caption{Number of groups and subhaloes identified by the FOF algorithm and \textsc{subfind} in \textsc{color} and \textsc{coco} at $z = 0 $.}
\label{halonums}
\end{table}

Since both \textsc{coco-warm} and its \textsc{cold} counterpart were
simulated using the same initial phases, we are able to match many
objects between the two simulations. This allows us to make also
object-by-object comparisons in addition to comparing just statistical
distributions of halo properties. In order to correctly match the
haloes we do the following: first, we take the 50 most-bound particles
from a \textsc{coco-warm} halo, and look for the \textsc{coco-cold}
halo in which there are at least 25 ($50\%$) of these particles. We
then confirm the match by repeating the same process, this time
starting with the \textsc{coco-cold} haloes, in decreasing order of
mass. This results in a bijective match between haloes in the two
simulations. Using this method, we are able to match
around 97\% of haloes with $M_{200} > 10^8~h^{-1}\,M_\odot$.

\begin{figure*}
\includegraphics[scale=0.72]{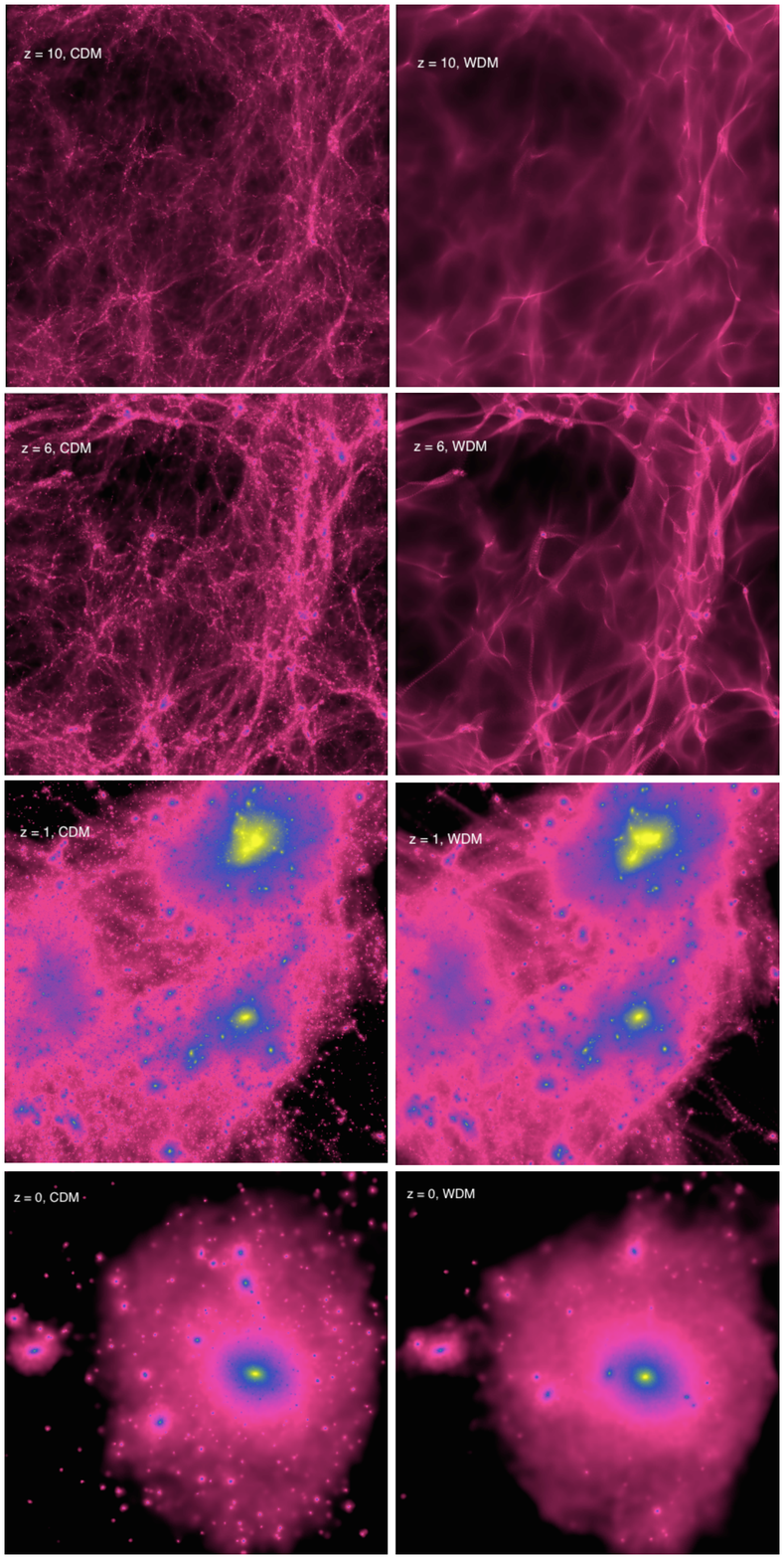}
\caption{Redshift evolution of the projected dark matter density in \textsc{coco-cold} (left) and the $3.3$ keV \textsc{coco-warm} Universe (right). From top to bottom, the top three panels show snapshots at $z=10, z=6, z=1$ of the projected mass density in cubes of side $2~h^{-1}\,\mathrm{Mpc}$, centred on the most massive group at $z = 0$. The bottom panels show zooms of a $5 \times 10^{10}~h^{-1}\,M_\odot$ halo at $z=0$ in a cube of side $150~h^{-1}\,\mathrm{kpc}$. The emergence of small haloes at early times is apparent in the CDM case, when the WDM distribution is much smoother. The formation of large haloes occurs at roughly the same time in the two simulations and the subsequent growth of these haloes is similar in the two cases. In the zoom shown in the bottom panel, the lack of substructure in the WDM case compared to its CDM counterpart is stark.}
\label{zsliceover}
\end{figure*}

\subsection{Spurious haloes and their removal}
\label{spurious}

Number counts of haloes and subhaloes are fundamental statistics of the
halo population, so the correct identification of haloes is of primary
importance. It has been known for some time
(\citealt{2007MNRAS.380...93W, 2013MNRAS.434.3337A,
  2014MNRAS.439..300L}) that in simulations in which the initial power
spectrum has a resolved cutoff, as is the case for \textsc{coco-warm}, 
small-scale structure is seeded in part by the discreteness of the particle set. In other words, 
a substructure finder will identify density peaks that have arisen not as a result of gravitational instabilities from
a cosmological perturbation. These
artificial fragments can often by identified ``by eye'' as they tend
to be regularly spaced along filaments of the mass distribution. They
produce a power-law-like upturn at small masses in the WDM mass
function. Since this is just a numerical (and resolution-dependent)
artefact of our WDM simulations, care must be taken to identify these
spurious haloes and, if appropriate, remove them from the halo
catalogue. While it is, in principle, possible to eliminate these
structures by increasing the resolution of the simulation, this is
computationally prohibitive: \cite{2007MNRAS.380...93W} have shown
that the mass at which spurious structures dominate the mass function
scales with the number of particles in the simulation, $N$, as $M
\propto N^{-1/3}$.

\cite{2014MNRAS.439..300L} developed an algorithm to identify spurious
clumps in WDM simulations. A large number of them can be removed by
performing a mass cut below a resolution-dependent scale, as suggested
by \cite{2007MNRAS.380...93W}:
\bq
M_{\mathrm{lim}} = 10.1 \bar{\rho}~ d~ k_{\mathrm{peak}}^{-2}\;,
\eq
where $d$ is the mean interparticle separation and $k_{\mathrm{peak}}$
is the spatial frequency at which the dimensionless power spectrum,
$\Delta^2(k)$, has its maximum. Applying this condition on its own
would also remove some genuine haloes that form below this scale.
\cite{2014MNRAS.439..300L} refined this criterion by also making a cut
on the basis of the shapes of the initial Lagrangian regions from
which WDM haloes form. They find that the spurious candidates tend to
have much more flattened configurations in their (unperturbed) initial positions than
genuine haloes, as judged from a CDM simulation. Defining the
sphericity, $s$, of haloes as the axis ratio, $c/a$, of the minor to
major axes in the diagonalised moment of inertia tensor of the initial
particle load, the sphericity cut is made such that $99\%$ of the CDM
haloes at that redshift lie above the threshold.

\begin{figure}
\includegraphics[width=0.5\textwidth]{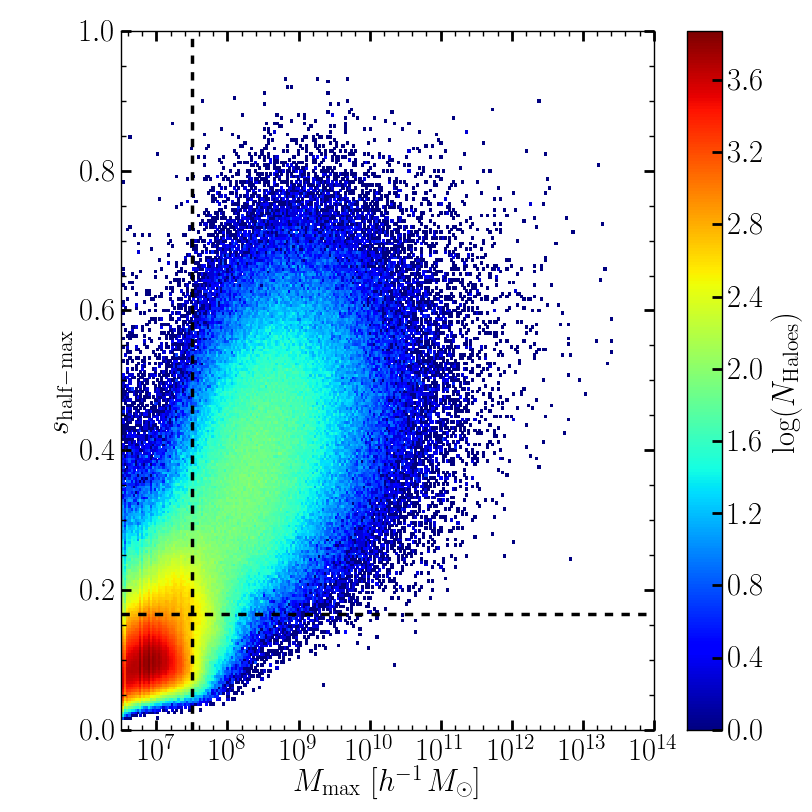}
\caption{Number density of haloes in the sphericity vs. maximum mass space in \textsc{coco-warm} at $z =0$. The dashed black lines show the cuts on sphericity and mass that we use to clean the halo catalogue. Rejected (spurious) candidates are those that fail the cuts in the manner described in the text.}
\label{spherdot}
\end{figure}

Following exactly the methodology of \cite{2014MNRAS.439..300L}, we
clean the \textsc{coco-warm} catalogue as follows: (1) remove all
(sub)haloes with $s_{\mathrm{half-max}} < 0.165$ \footnote{The
  criterion $s_{\mathrm{half-max}} < 0.165$ is appropriate for haloes
  identified at $z=0$; for higher redshifts, one needs to determine
  the $1\%$ sphericity cut at {\em that} redshift.}, irrespective
of mass; (2) for those that pass (1), remove (sub)haloes with
$M_{\mathrm{max}} < 0.5 M_{\mathrm{lim}}$. Here, $M_{\mathrm{max}}$ is
the maximum mass attained by a (sub)halo during its evolution, and
$s_{\mathrm{half-max}}$ is the sphericity ($= c/a$) of the (sub)halo
at the half-maximum mass snapshot. This is chosen so as to identify a
(sub)halo at a time well before it falls into a larger host, when its
particles are subject to tidal stripping.  The factor of 0.5 in
condition (2) is calibrated by matching between resolutions in the
\textsc{aquarius} simulations (see \citealt{2014MNRAS.439..300L} for
details). Having done so, we find that over $91\%$ of the (FOF) haloes
formed in \textsc{coco-warm} are in fact spurious, and are rejected
from the halo catalogue when computing properties like mass
functions. The elements of this section are summarised in
Fig.~\ref{spherdot}.

\section{Results}
\label{Results}

In both cold and warm models, dark matter haloes assemble in a
hierarchical way, acquiring mass by merging with other haloes and by
smoothly accreting ambient mass
\citep[\eg][]{1974ApJ...187..425P,Frenk1985Natur,MergersLaceyCole1993,Wechsler2002}. In
this section, we focus on global halo properties such as formation
times, abundance and internal structure. We make a direct comparison
between our cold and warm dark matter models. On scales much larger
than the WDM suppression scale in the initial power spectrum, we expect the
properties of haloes to be very similar in the two cases, but
differences should become increasingly important at $\sim 2 \times 10^9~h^{-1}\,M_\odot$
and below.

\subsection{Redshift of formation}
\label{zform}

The absence of primordial perturbations below the cutoff scale in the WDM
power spectrum induces differences in the
formation times of the smallest haloes. We can visualise these
differences directly by examining the images displayed in
Fig.~\ref{zsliceover}. At early times, the projected density in
\textsc{coco-warm} (right panels) is visibly smoother than the
equivalent projection in \textsc{coco-cold} (left panels), which has a
``grainier'' appearance owing to the very large number of haloes below
$\sim 10^9~h^{-1}\,M_\odot$ that form in this case, well before the first
objects have collapsed in \textsc{coco-warm}. Thus, the onset of the
structure formation process in this simulation is delayed relative to
its CDM counterpart.

In order to quantify the different halo formation epochs in
\textsc{coco-warm} and \textsc{coco-cold}, we trace the evolution of
each FOF group through its merger tree, and define the redshift of
formation as the first time when the mass of the most massive
progenitor exceeds half the final FOF mass:
$M\left(z_{\mathrm{form}}\right) = M\left(z = 0\right) /2$
(e.g. \citealt{2006MNRAS.367.1039H,2007MNRAS.381.1450N}). Other
definitions of halo formation time also exist in the literature
(e.g. \citealt{1996ApJ...462..563N,1997ApJ...490..493N}), which should
be borne in mind when making comparisons.

\begin{figure}
\includegraphics[width=0.5\textwidth]{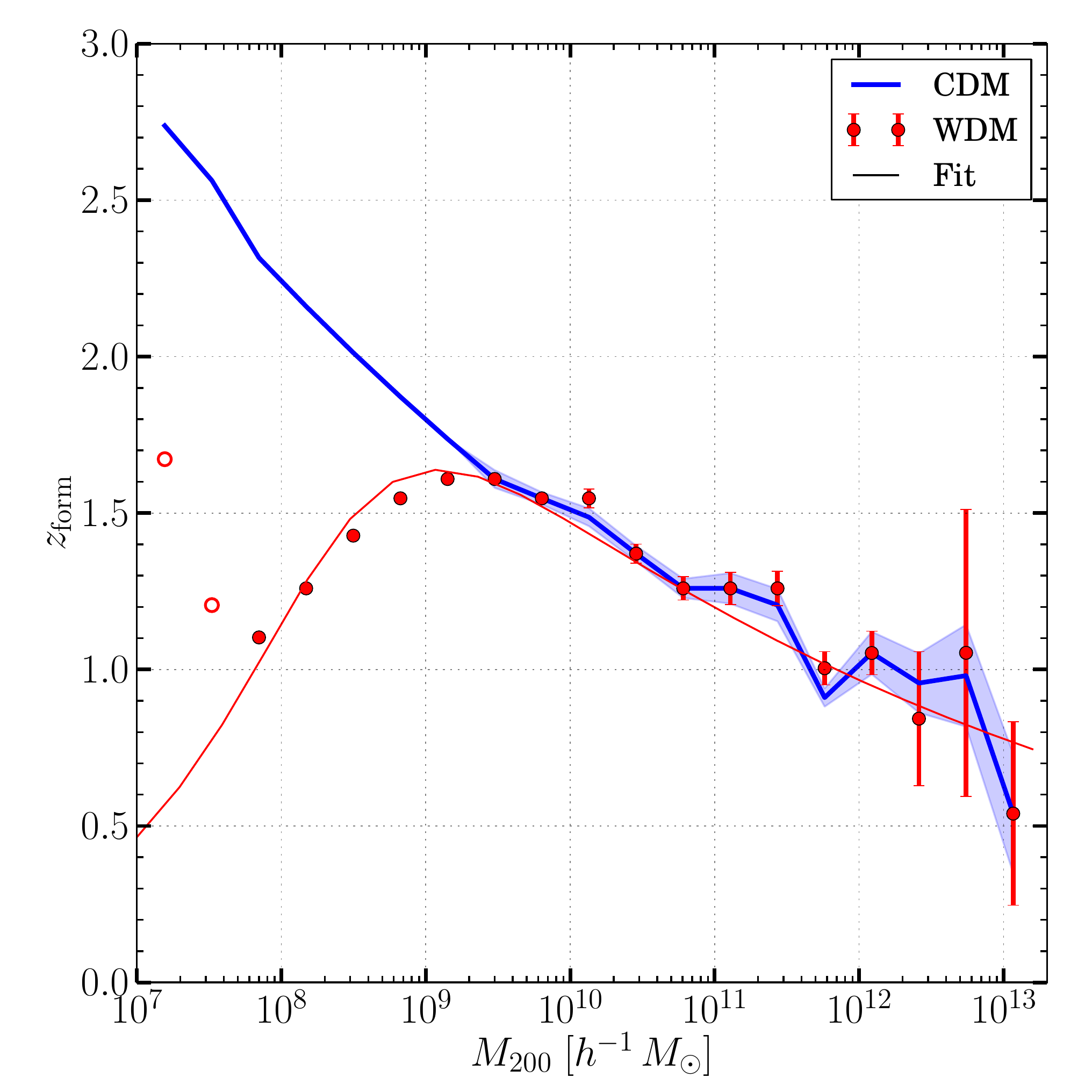}
\caption{The median redshift of formation of all FOF groups in
  \textsc{coco-warm} and \textsc{coco-cold}, as a function of the halo
  mass, $M_{200}$. The redshift $z_{\mathrm{form}}$ is defined in the
  text. The error bars / shaded region represent the bootstrapped
  errors on the median in each mass bin in \textsc{coco-warm} and
  \textsc{coco-cold} respectively.  As expected, there is good
  agreement at the high-mass end, whereas the differences between CDM
  and WDM become apparent below $\sim 2 \times
  10^9~h^{-1}\,M_\odot$.  The thin red line is a fit to the
  \textsc{coco-warm} redshift of formation, using Eq.~\ref{fitzform}.}
\label{z_form}
\end{figure}

The result, for {\it all} haloes in \textsc{coco-warm} (including
spurious objects) and \textsc{coco-cold} is shown in
Fig.~\ref{z_form}. The formation redshifts of haloes of mass
$M_{\mathrm{200}} \gtrsim 2 \times 10^9~h^{-1}~M_\odot$, are very similar in
\textsc{coco-warm} and \textsc{coco-cold}, as expected. The difference
between the two begins to manifest below a mass of $M_{\mathrm{200}}
\sim 2 \times 10^9~h^{-1}~M_\odot$, an order of magnitude above
the half-mode mass scale for a $3.3$ keV WDM particle
(c.f. Section~\ref{SimParams}). For these smaller haloes,
$z_{\mathrm{form}}$ is lower for WDM than CDM. The sudden upturn in
the WDM $z_{\mathrm{form}}$ for $M_{\mathrm{200}} <
10^8~h^{-1}~M_\odot$ (shown in the open red circles) is a signature of
the spurious haloes described in Section~\ref{spurious}. From here on,
we will exclude these spurious haloes and only show results from the
cleaned \textsc{coco-warm} sample. The difference in formation times
is a subject we will revisit when comparing the concentration-mass
relations of WDM and CDM in Section~\ref{Conc}. Note that in this
figure, we include all haloes, and not necessarily matched between CDM
and WDM, which is why the medians at the largest mass bins are not
exactly identical.

We find that the delay in the formation time of \textsc{coco-warm}
haloes of a given mass, relative to \textsc{coco-cold}, is
well described by the fitting function:
\bq \label{fitzform}
\frac{z_{\mathrm{form}}^{\mathrm{WDM}}}{z_{\mathrm{form}}^{\mathrm{CDM}}} = \left(1 + a\frac{M_{\mathrm{hm}}}{M_{200}} \right)^{-b}\;,
\eq
where $M_{\mathrm{hm}}$ is the half-mode mass introduced in
Section~\ref{SimParams}, $a = 1.23$ and $b =0.56$. This fit is shown
as the thin red line in Fig.~\ref{z_form}.

\subsection{Differential halo mass functions}
\label{HMF}

Counting the number of dark matter haloes as a function of their mass
is one of the simplest and most important population statistics that
one can use to distinguish between WDM and CDM models, since fewer
haloes will form in the former close to the half-mode mass. 

\begin{figure}
\includegraphics[width=0.5\textwidth]{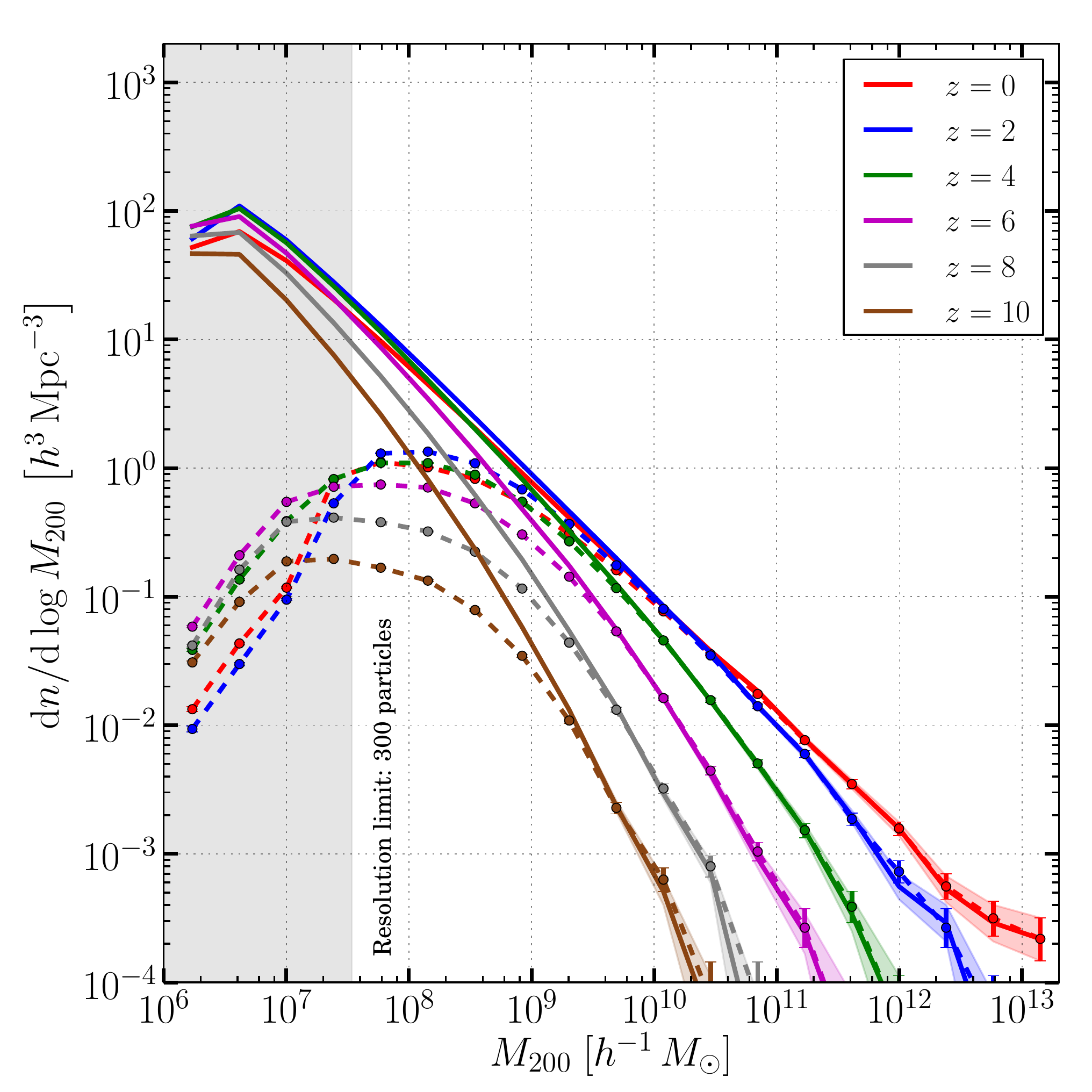}
\caption{The redshift evolution of the halo mass function in
  \textsc{coco-cold} and \textsc{coco-warm}.  The solid lines show the
  CDM results, with the shaded regions representing the associated
  $1\sigma$ Poisson errors. The dashed lines with error bars
  represent the equivalent relation from \textsc{coco-warm}, with
  spurious haloes removed. The different colours show results for a
  selection of redshifts, as indicated in the legend. The grey shaded
  region corresponds to haloes with fewer than 300 particles within
  $r_{200}$.}
\label{HMFEvol}
\end{figure}

In Fig.~\ref{HMFEvol}, we show the build-up of the halo population as
a function of redshift in \textsc{coco-cold} (solid lines) and
\textsc{coco-warm} (dashed lines). The shaded regions and error bars
represent the Poisson uncertainty in both cases. Spurious haloes have
been omitted from the WDM differential halo mass function (dHMF) at
each redshift, using the methodology outlined in
Section~\ref{spurious}. The edge of the grey region marks the nominal
resolution limit of our simulation which corresponds to a halo with at
least 300 particles within $r_{200}$ ($M_{200} \sim 3.4 \times
10^{7}~h^{-1}\,M_\odot$). This 300-particle limit was derived by
comparing the mass function of \textsc{coco-cold} with that of its
lower-resolution counterpart \textsc{color-cold}. Below this limit,
the results of the simulations become increasingly unreliable. The
results at high masses are noisy because of the small number of
high-mass haloes formed in the relatively small volume of our
simulations.

The general trend across redshifts is similar: for haloes with
$M_{\mathrm{200}} > 2 \times 10^{9}~h^{-1}\,M_\odot$, the dHMF in
\textsc{coco-warm} and \textsc{coco-cold} are almost identical. The
abundance of haloes below this mass scale is strongly suppressed in
\textsc{coco-warm}, to the extent that, at $z=10$, there are 5 times
fewer~$\sim10^8~h^{-1}\,M_\odot$ haloes than in
\textsc{coco-cold}. The delayed non-linear structure formation below
$\sim 2 \times 10^9~h^{-1}\,M_\odot$ can also be seen from the fact that there are as many
haloes with 
$M_{200}~=~10^8~h^{-1}\,M_\odot$  in \textsc{coco-warm} at $z =
10$, as there are haloes with $M_{200} = 6 \times 10^8~h^{-1}\,M_\odot$ in
\textsc{coco-cold} at that redshift.

Within the CDM paradigm, there are a number of analytic predictions for the
differential halo mass function (dHMF), notably the Press-Schechter
formula (\citealt{1974ApJ...187..425P,
  1991ApJ...379..440B,MergersLaceyCole1993}), and the ellipsoidal
collapse model (ST; \citealt{1999MNRAS.308..119S}, although this model
is not fully analytic since it is tuned to numerical simulations). The
dHMF is given by:
\bq \frac{\mathrm{d}n}{\mathrm{d}\log M} =
\frac{\bar{\rho}}{M}f\left(\nu\right)\left|\frac{\mathrm{d\log
      \sigma^{-1}}}{\mathrm{d}\log M}\right|\;, \eq
where $f(\nu)$ is the so-called {\it halo multiplicity function} and
for hierarchical cosmologies has a universal form \citep[see
\eg][]{MFJenkins2001,MFReed2007,TinkerMF2008,MXXLAngulo2012}. In the
ST formalism, it is approximated by:
\bq
f\left(\nu\right) = A \sqrt{\frac{2q\nu} {\upi}} \left[1 + \left(q\nu\right)^{-p}\right] e^{-q\nu/2}.
\eq
Here, $\nu \equiv \delta_c^2 (z)/ \sigma^2(M), A = 0.3222, q = 0.707$
and $p = 0.3$. In linear theory, $\delta_c (z) \equiv 1.686/D(z)$,
where $D(z)$ is the linear growth rate of perturbations. The value of
$\delta_c$ is appropriate for the Einstein-de Sitter model, but
differs slightly in $\Lambda$CDM due to a weak dependence on
$\Omega_m(z)$.  Finally, $\sigma^2(M)$ is the variance in the mass
density field on mass scale, $M$, given by:
\bq \label{sig}
\sigma^2(M) = \int \frac{\textrm{d}k}{k} \Delta^2(k) \tilde{W}^2(k,M)\,.
\eq 
Here, $\tilde{W}(k,M)$ is the Fourier transform of a window function containing mass $M$, and $\Delta^2(k)$ is the dimensionless power spectrum as defined in Fig.~\ref{matterps}. 

\begin{figure}
\includegraphics[width=0.5\textwidth]{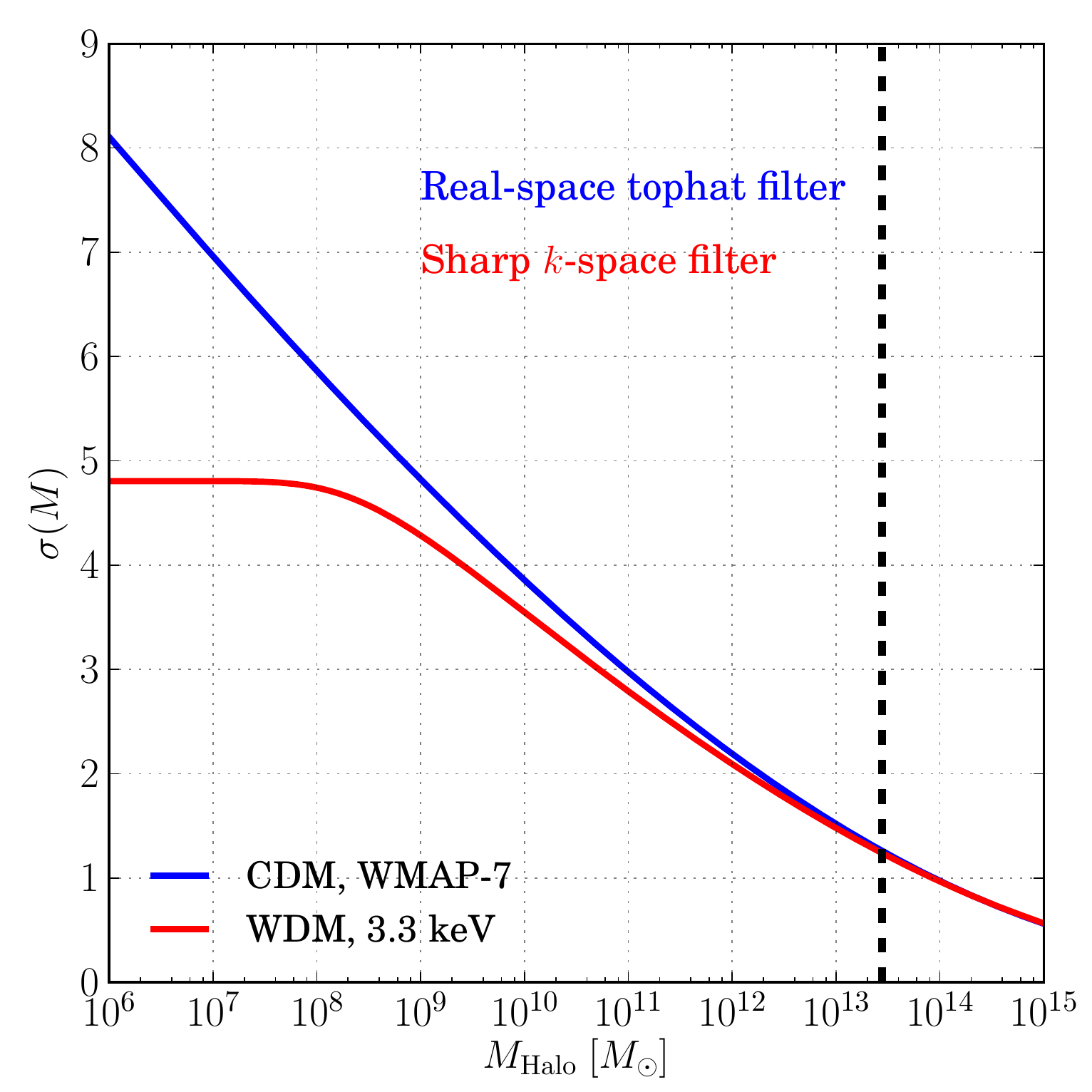}
\caption{The fractional variance of the density field, $\sigma(M)$,
  calculated in Eq.~\ref{sig} using a top-hat filter in real-space for
  CDM, and a sharp $k$-space filter for WDM. The flattening of the
  relation below $10^8~h^{-1}~M_\odot$ is due to the suppression of
  power below these scales in WDM, relative to CDM. The dashed line
  indicates the upper limit to the halo masses formed in our
  volume-limited simulations.}
\label{sigmaM}
\end{figure}

In the Press-Schechter and Sheth-Tormen formalisms, the {\it rms}
fluctuation amplitude, $\sigma^2(M)$, is assumed to be a monotonically increasing
function of $M$. This is no longer true for the truncated power
spectrum of WDM, so care must be taken when choosing an appropriate
window function. In the CDM, $W(k,M)$ is usually chosen to be the
real-space spherical top-hat function, a choice that results in an
excellent match to the dHMF in cosmological $N$-body simulations. The
same for WDM predicts an excess of low-mass haloes compared to
simulations (\citealt{2001ApJ...558..482B, 2012MNRAS.421.2384M,
  2012MNRAS.424..684S}, but see also
\citealt{2013MNRAS.433.1573S}). This problem was solved by
\cite{2013MNRAS.428.1774B}, who generalised the (extended)
Press-Schechter formalism by using the correct solution for the
excursion set barrier first-crossing distribution in WDM
models. Rather than the top-hat real-space window function, they used
a sharp $k$-space filter for WDM, so that the variance, $\sigma(M)$,
remains flat up to the half-mode mass and then declines with increasing
mass (see Fig.~\ref{sigmaM}). In this formalism the smoothing scale,
$R$, is defined as:
\bq
R = \frac{a}{k_s}\;,
\eq
where $k_s = 2\upi \kappa / \alpha$, $\alpha$ as defined in Eq.~\ref{alphaMass}, $\kappa = 0.361$ and $a =
2.5$. \cite{2013MNRAS.428.1774B} choose the free parameters such that
the theoretical mass function turns over at the same scale as the halo
mass function from simulations. This choice of parameters should be
applicable to all thermal WDM models, since the effect of the WDM
suppression is captured in the value of $\alpha$
(Eq.~\ref{alphaMass}).

\begin{figure}
\includegraphics[width=0.5\textwidth]{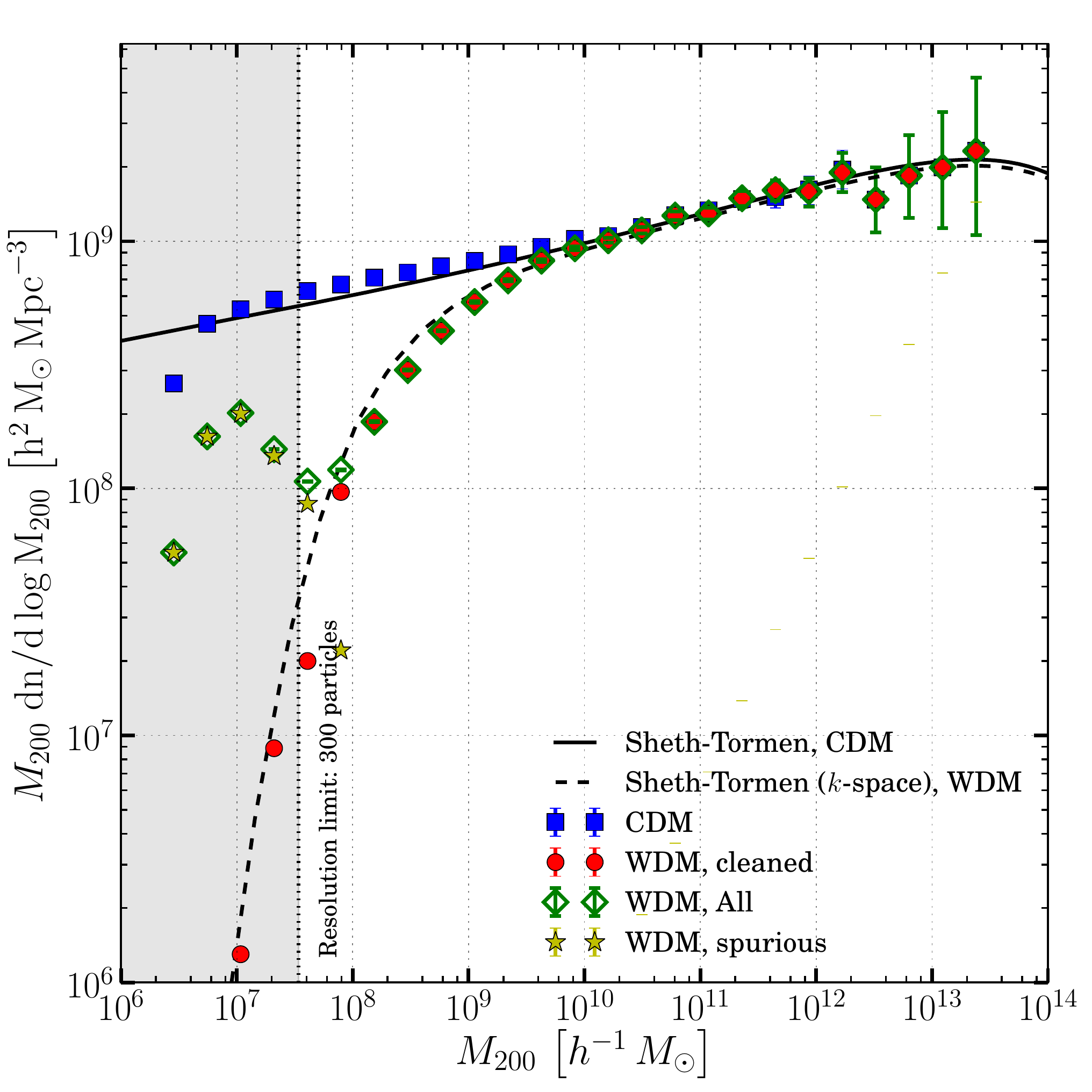}
\caption{Differential halo mass functions from the \textsc{coco-warm}
  and \textsc{coco-cold} simulations, compared to the predictions of
  the ellipsoidal collapse formalism of
  \citet{1999MNRAS.308..119S}. The solid lines show the predictions of
  the standard formalism applied to CDM; the dashed lines show the
  predictions of the modified, sharp $k$-space filter of
  \citet{2013MNRAS.428.1774B}. The symbols represent results from our
  simulations as denoted in the legend: blue squares for
  \textsc{coco-cold}, green diamonds for all \textsc{coco-warm} FOF
  haloes, red circles for the genuine haloes and yellow stars for
  spurious haloes. The grey shaded region corresponds to haloes with fewer
  than 300 particles within $r_{200}$. }
\label{sheth}
\end{figure}

In Fig.~\ref{sheth}, we compare the $z =0$ dHMF for \textsc{coco-cold} (blue
squares), the full \textsc{coco-warm} (genuine and spurious objects;
green diamonds), the spurious \textsc{coco-warm} objects only (yellow
stars) and the genuine \textsc{coco-warm} haloes only (red
circles). 

The solid and dashed black lines in Fig.~\ref{sheth} show the ST
predictions for the mass functions in CDM and WDM respectively.  For
$M_{200} > 2 \times 10^{9}~h^{-1}\,M_\odot$, the mass functions for CDM and WDM
trace one another exactly, as expected. Below this mass, the WDM mass
function begins to peel off from the CDM case, reaching half the CDM
amplitude at $M_{200} \sim 2 \times 10^{8}~h^{-1}\,M_\odot$. This
agrees with the half-mode mass scale, $M_{\mathrm{hm}}$, introduced in
Section~\ref{SimParams}. The raw WDM mass function (green diamonds) is
entirely dominated by the spurious objects (yellow stars) below $\sim
4 \times 10^7~h^{-1}\,M_\odot$, where the mass function shows an
artificial upturn. On the other hand, the cleaned WDM sample,
represented by the red circles, continues to fall off smoothly from
the regime free of artificial haloes. The feature at
$\sim 2 \times 10^7~h^{-1}\,M_\odot$ could be related to the cut,
$M_{\mathrm{max}} = 3.2 \times 10^7~h^{-1}\,M_\odot$, applied as part
of the cleaning procedure (Section~\ref{spurious}), but, in any case,
this is very close to the resolution limit which is also the mass
scale at which the spurious haloes begin to dominate the mass
function.

The main conclusion to be drawn from Fig.~\ref{sheth} is that above
the resolution limit, the modified ellipsoidal collapse model
reproduces the WDM mass function remarkably accurately, over nearly 6
orders of magnitude in mass.

\subsection{Halo density and mass profiles}

\label{density}

\begin{figure*}
\includegraphics[scale=0.38]{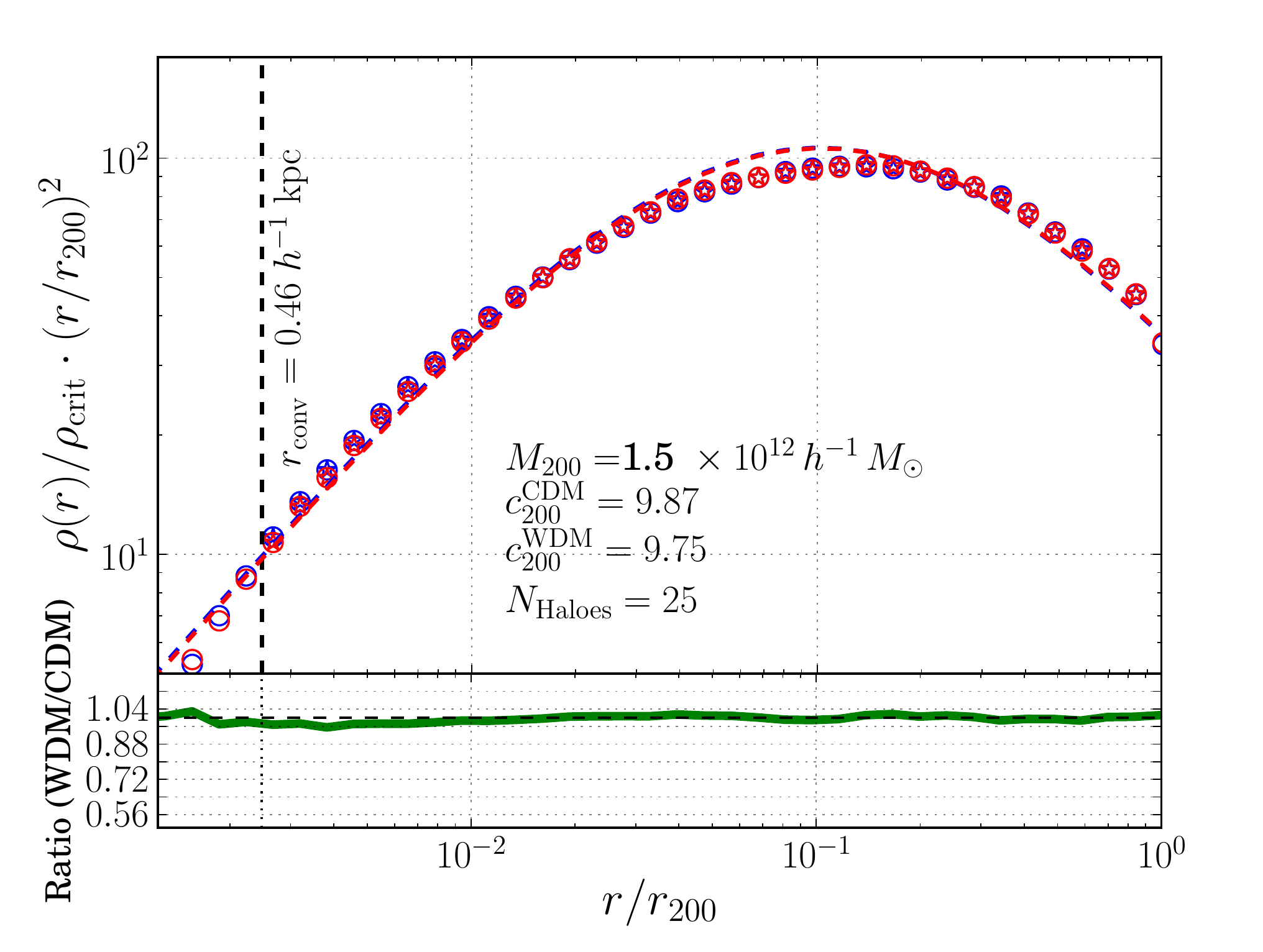}
\includegraphics[scale=0.38]{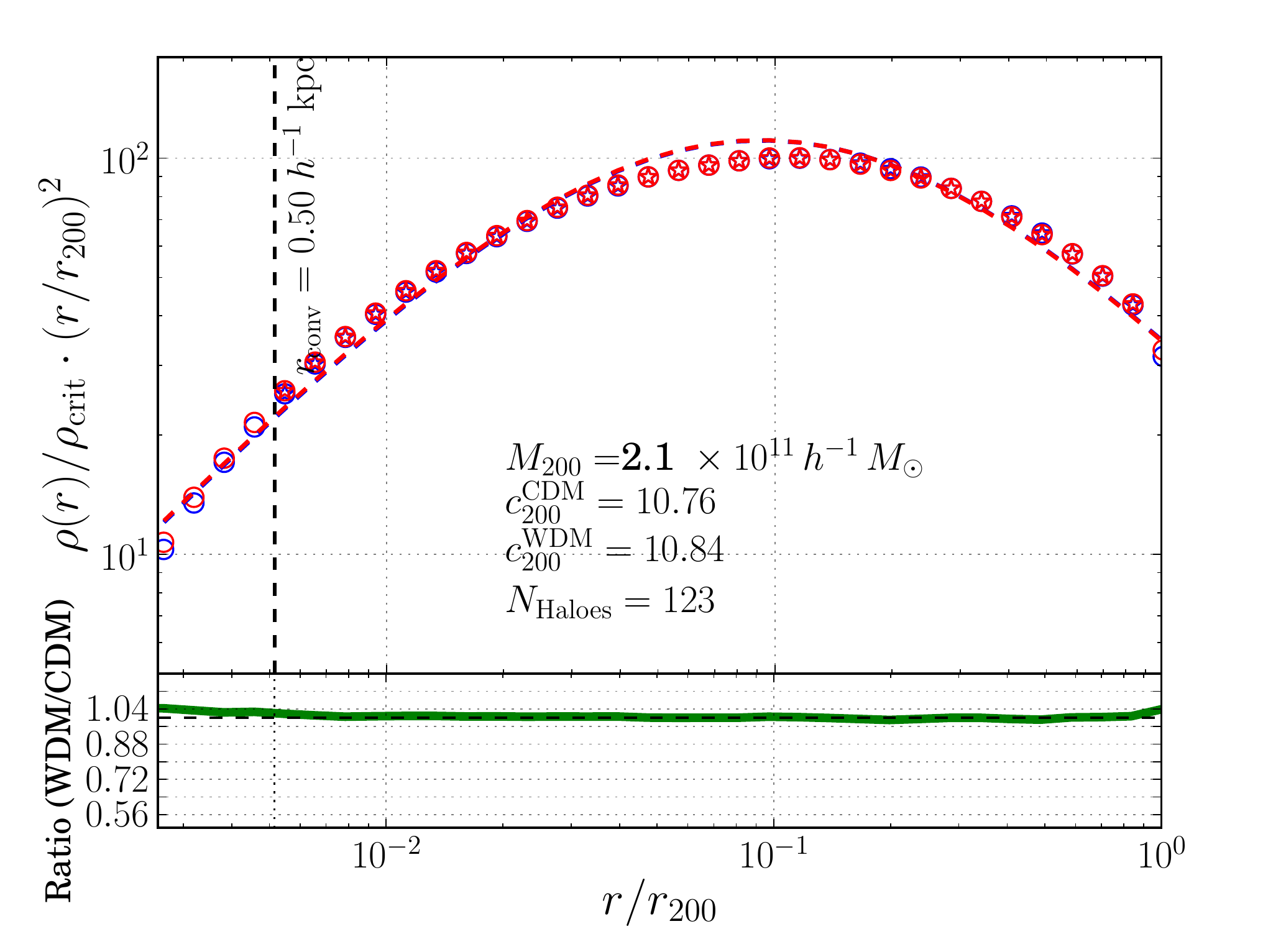} \\ 
\includegraphics[scale=0.38]{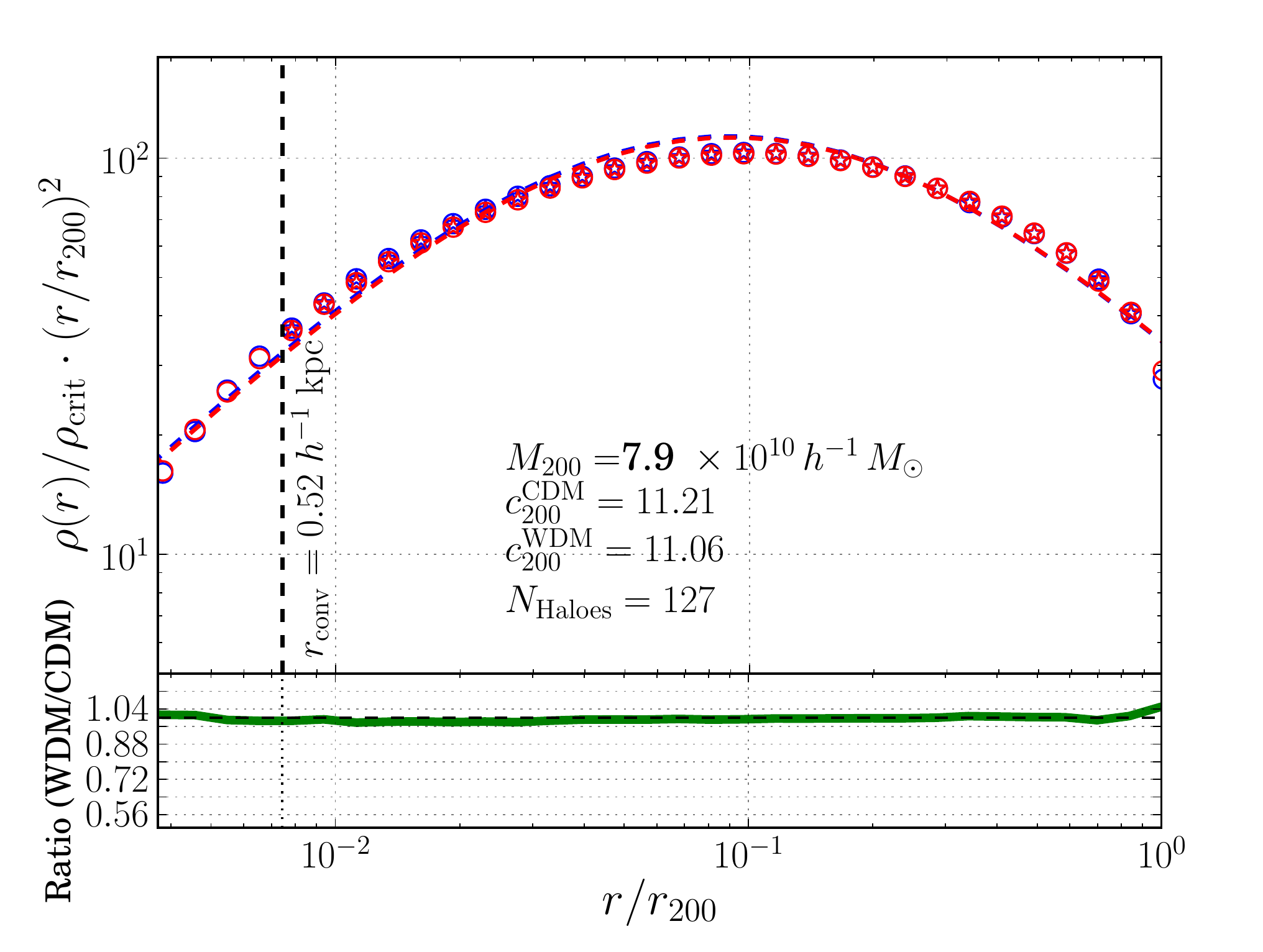}
\includegraphics[scale=0.38]{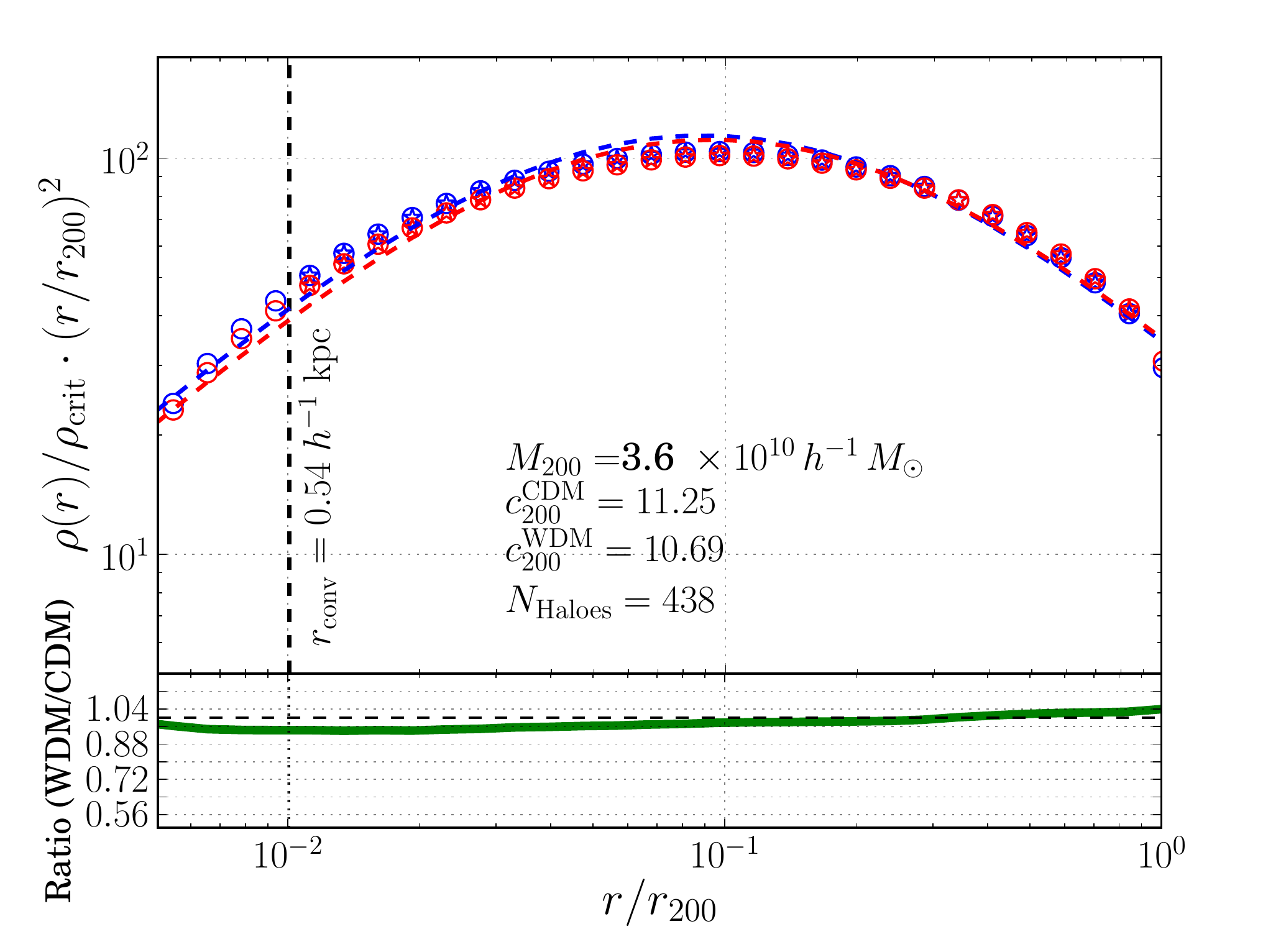}\\
\includegraphics[scale=0.38]{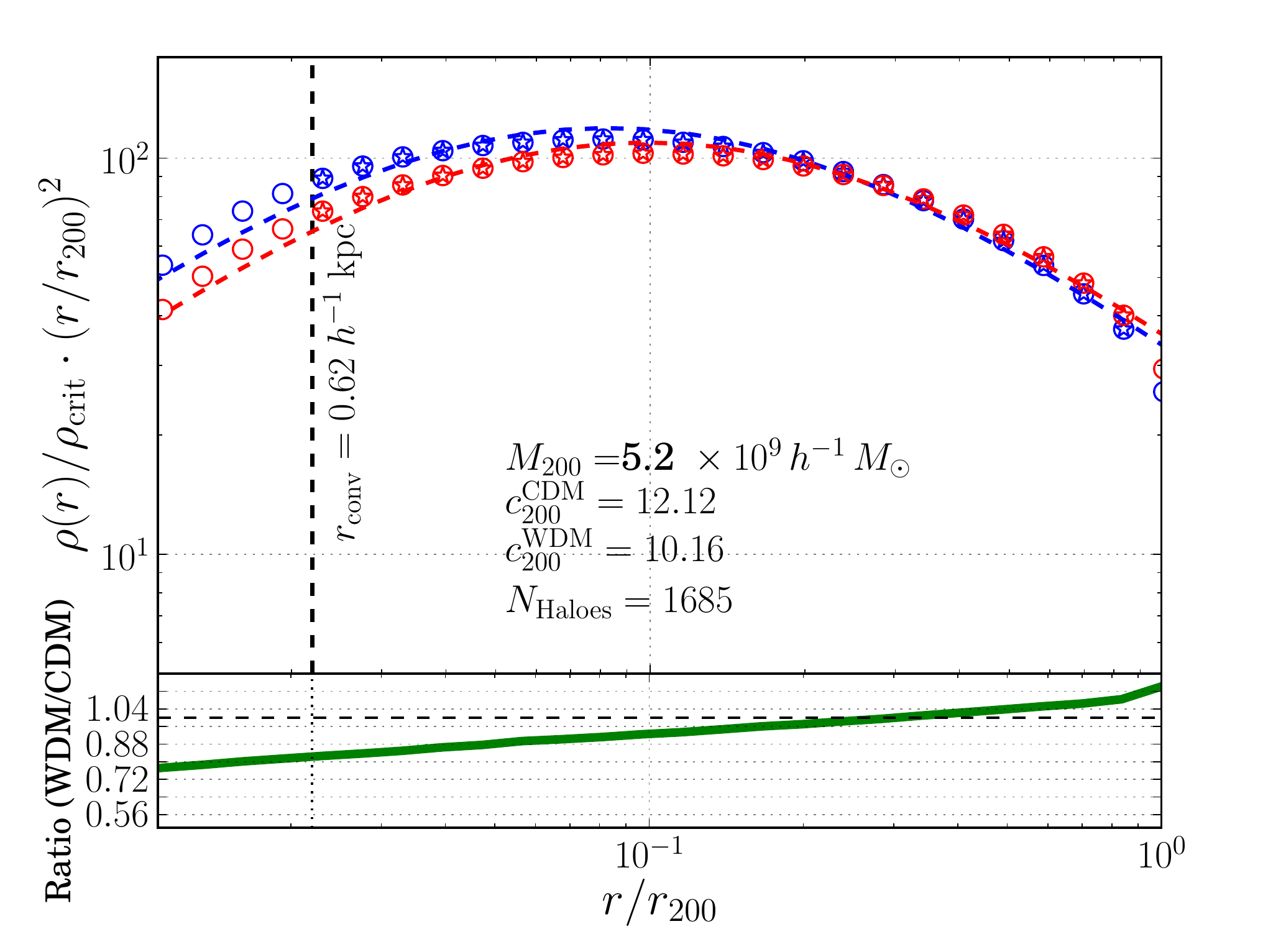}
\includegraphics[scale=0.38]{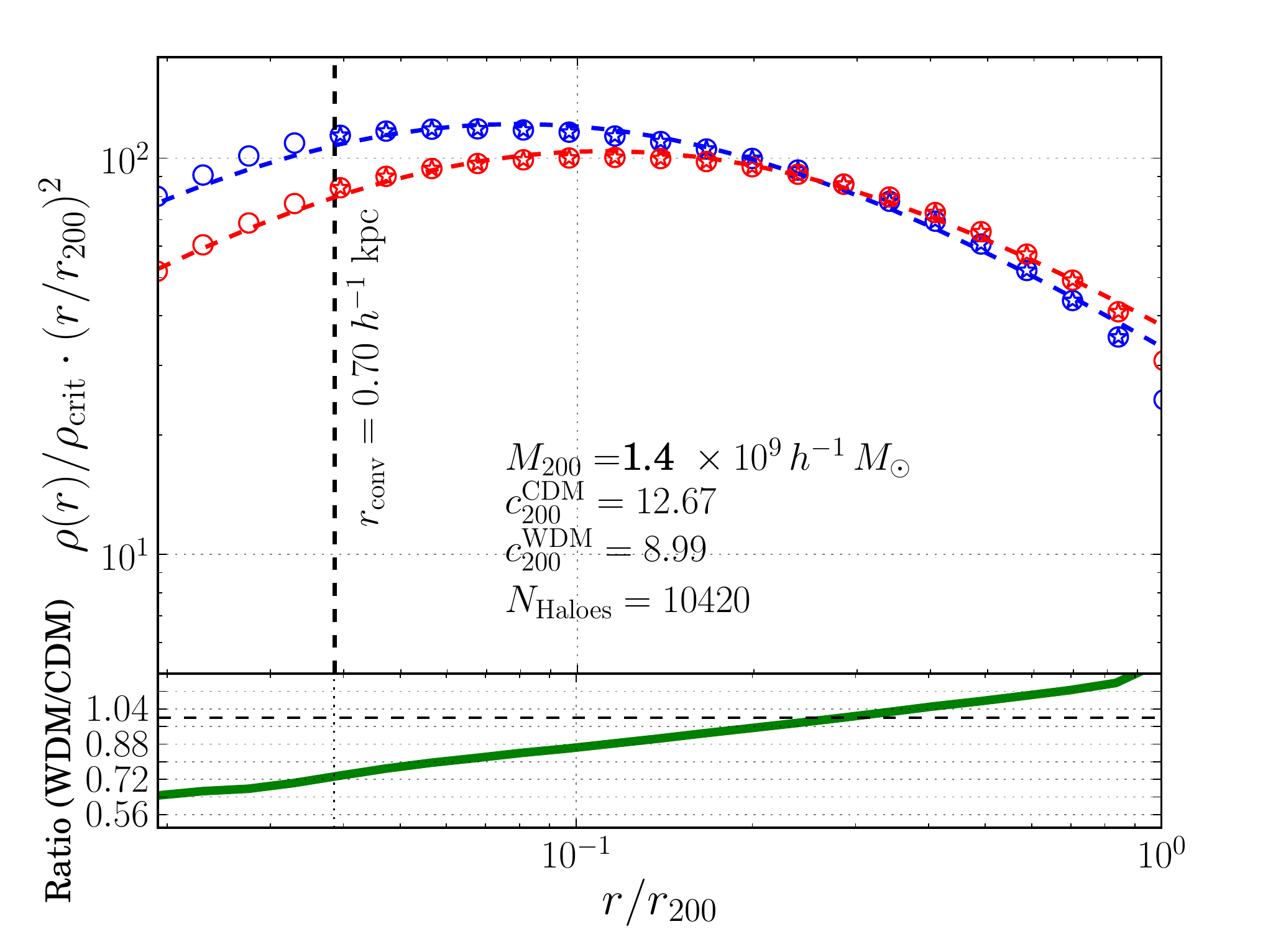}\\
\caption{Stacked spherically-averaged density profiles in
  \textsc{coco-warm} (red) and \textsc{coco-cold} (blue). For each
  mass bin we compare the profiles of only relaxed, matched haloes in
  the two simulations; the number in each bin is indicated in each
  subpanel. The vertical dashed line represents the convergence
  radius, $r_{\mathrm{conv}}$, and filled symbols indicate the range
  of the profile above this limit, whereas open symbols denote the
  radial range below it. The dashed red and blue lines are NFW fits to
  the WDM and CDM profiles respectively. Note that the density
  profiles have been scaled by $(r/r_{200})^2$ so as to reduce the
  dynamic range on the vertical axis. The bottom panels show the ratio
  of the WDM and CDM densities in each bin.}
\label{stackden}
\end{figure*}

\begin{figure*}
\includegraphics[scale=0.38]{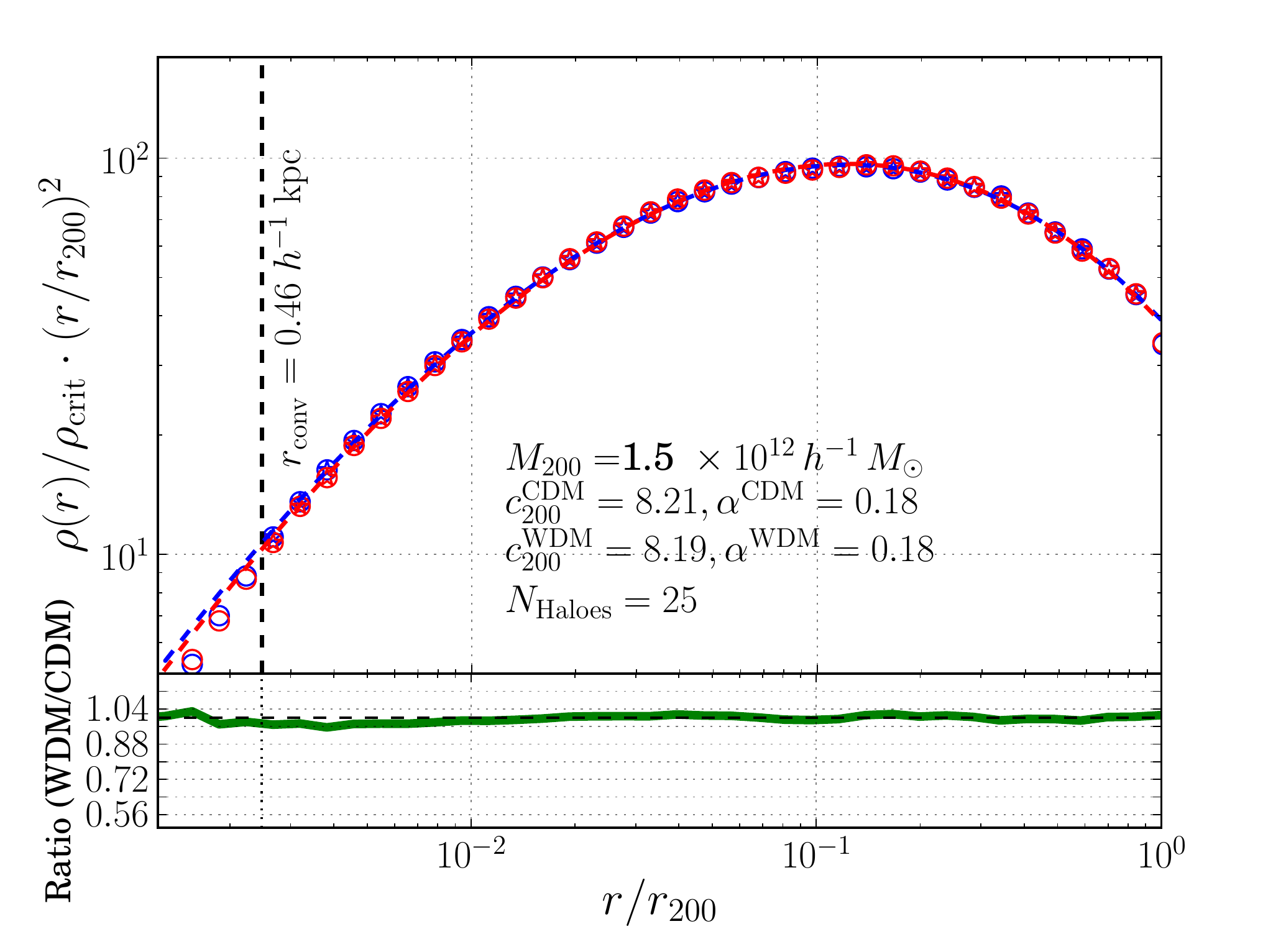}
\includegraphics[scale=0.38]{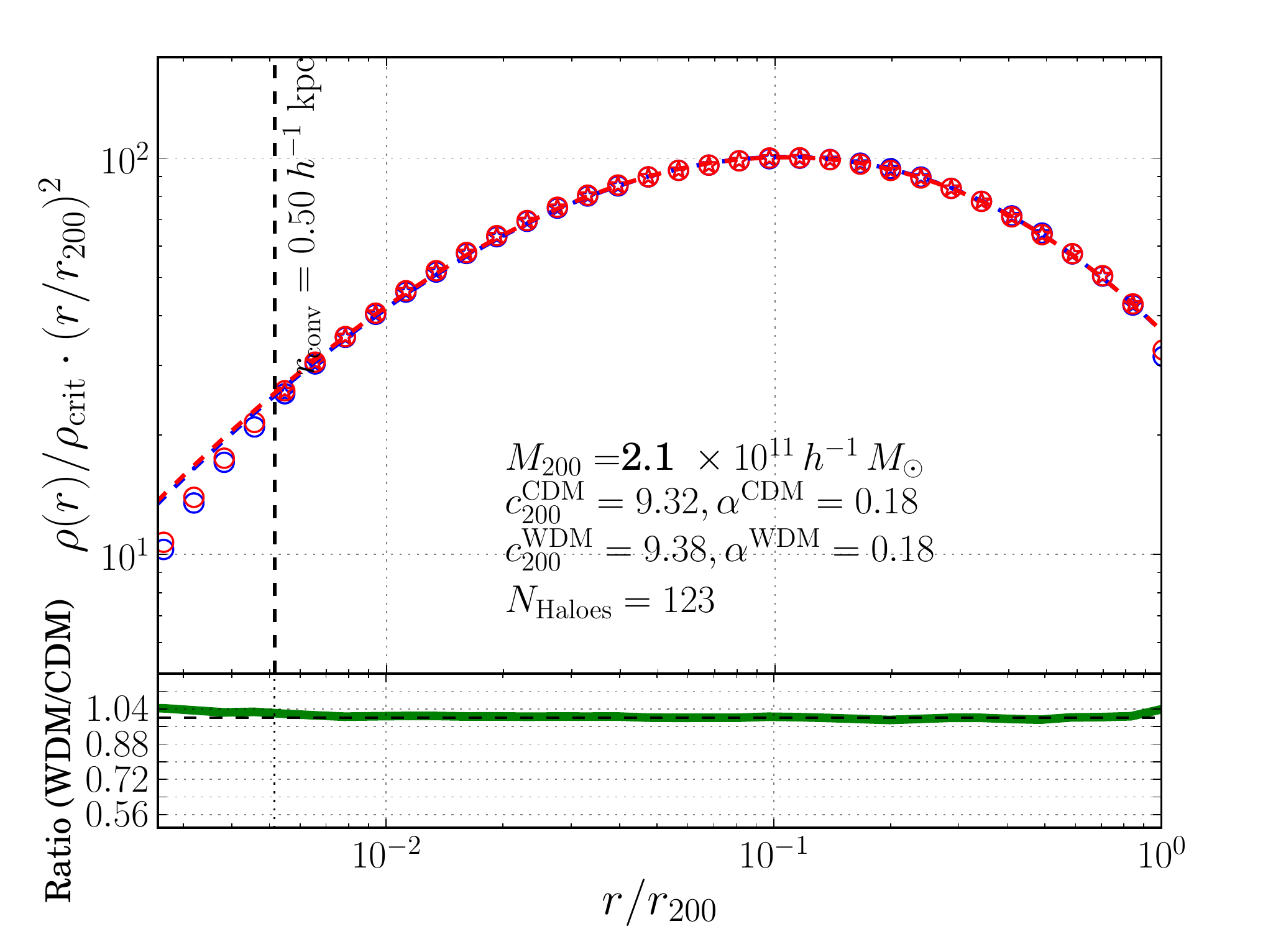} \\ 
\includegraphics[scale=0.38]{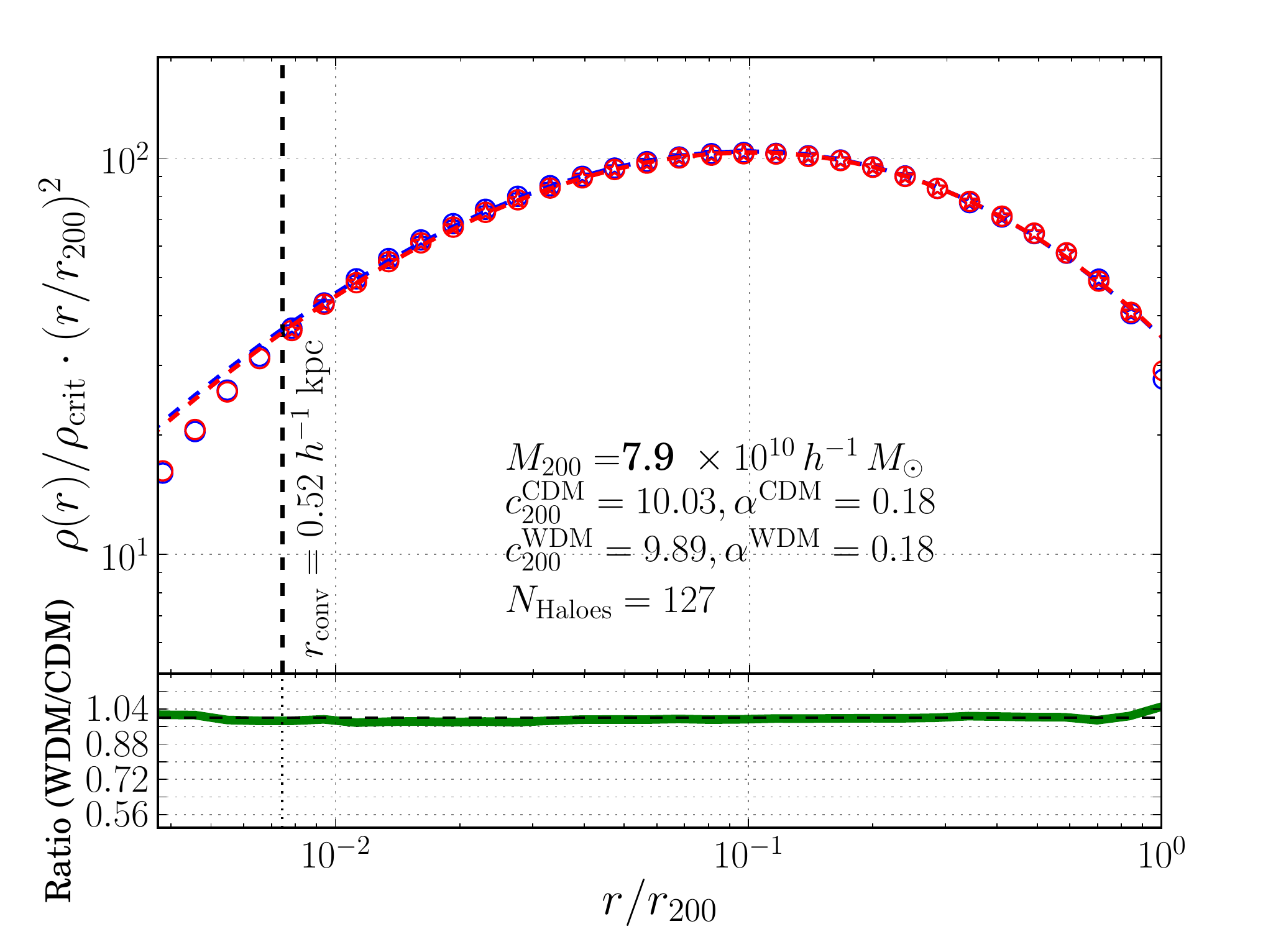}
\includegraphics[scale=0.38]{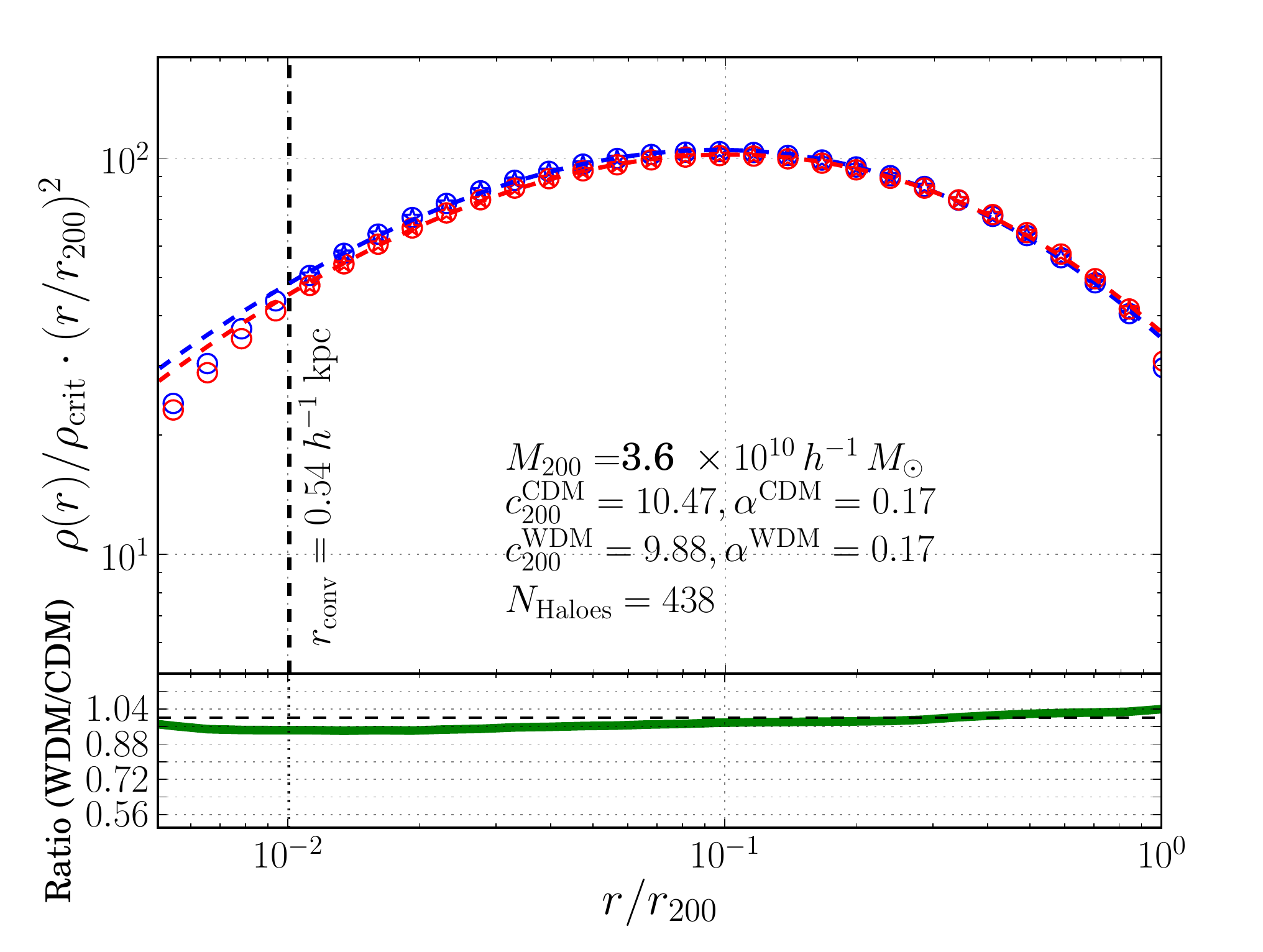}\\
\includegraphics[scale=0.38]{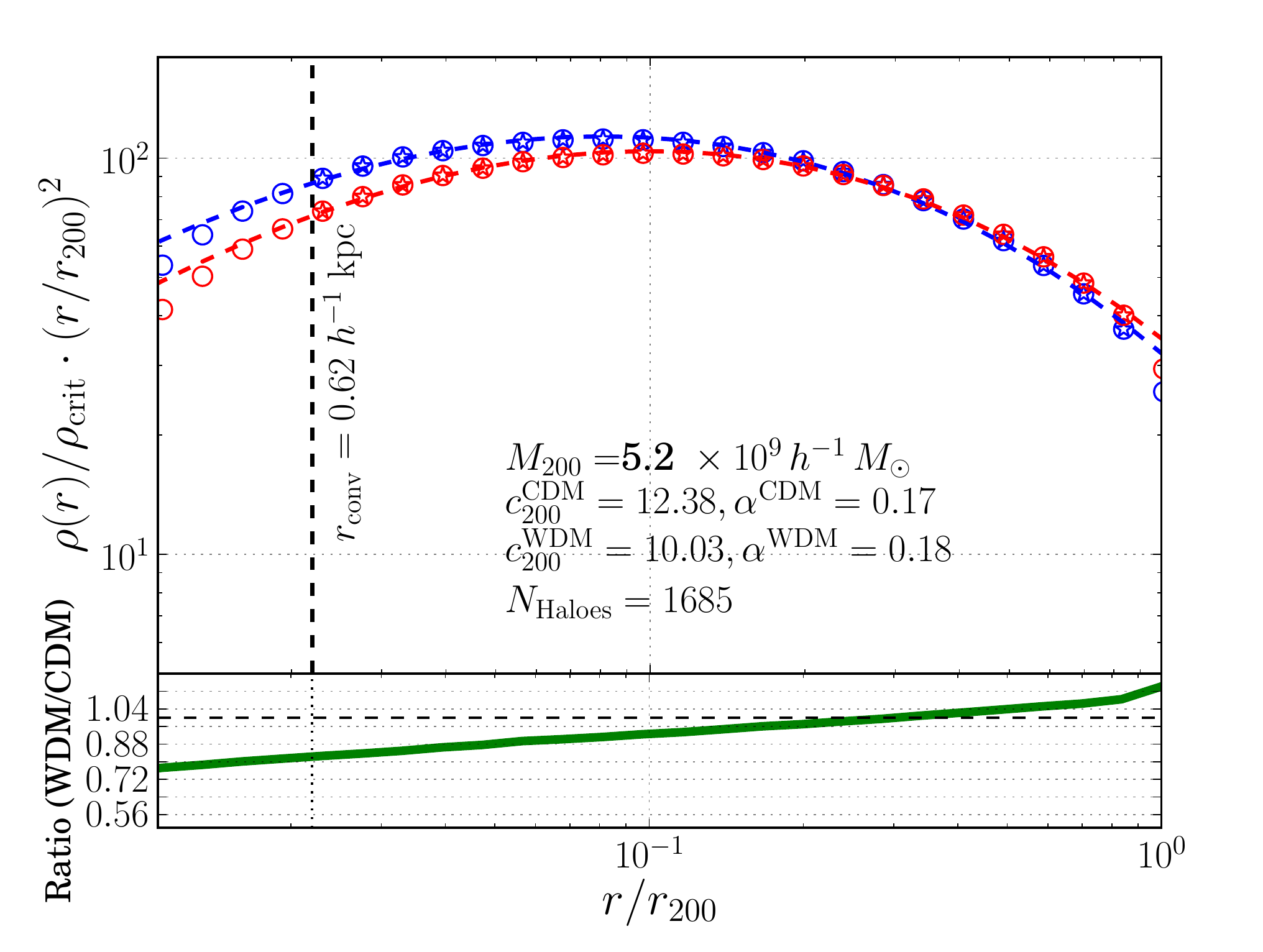}
\includegraphics[scale=0.38]{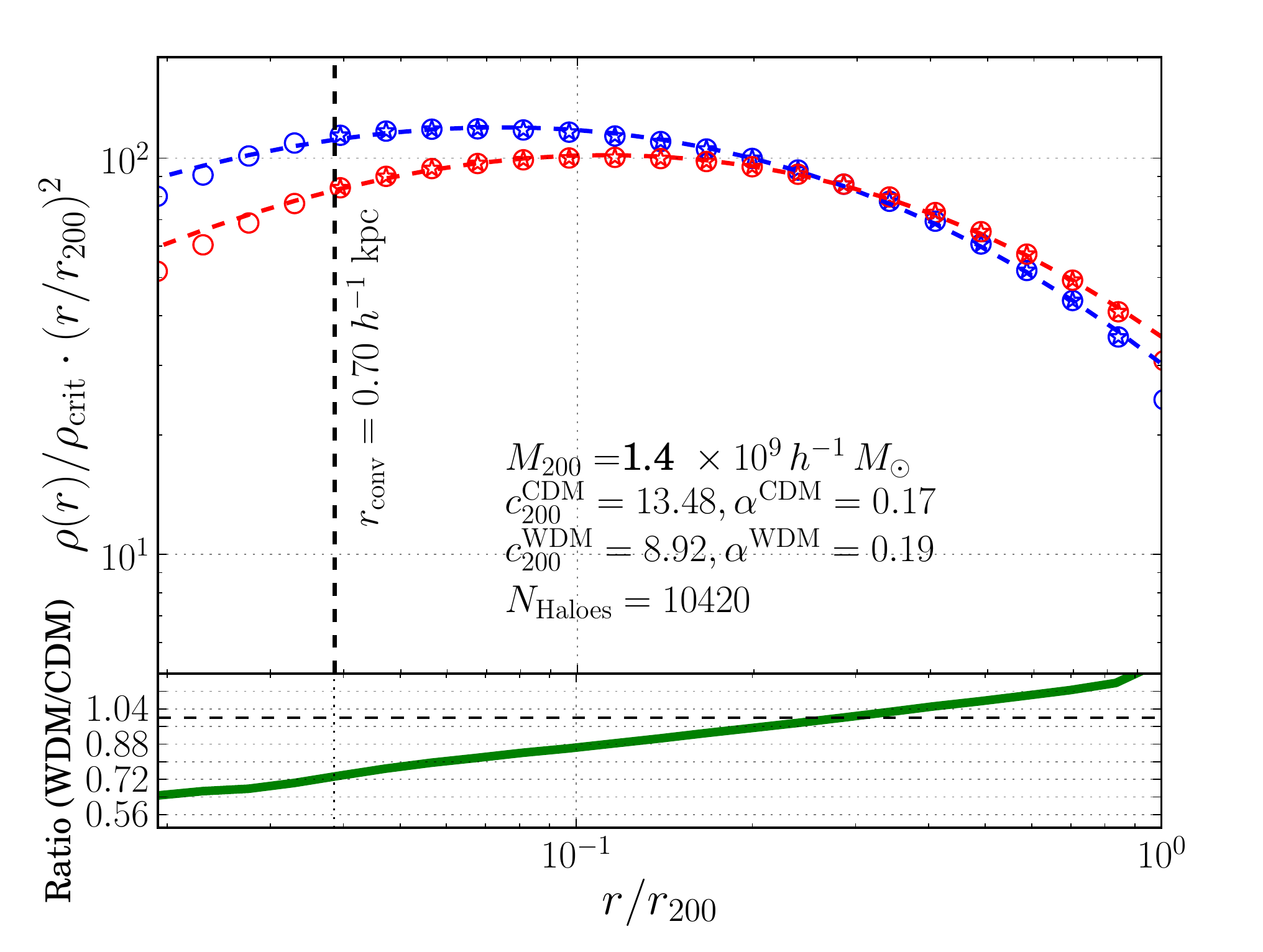}\\
\caption{Same as Fig.~\ref{stackden}, but with Einasto fits to the
  \textsc{coco-warm} and \textsc{coco-cold} density profiles.}
\label{stackdenEin}
\end{figure*}

Spherically-averaged radial density profiles provide the simplest and
most direct descriptor of halo structure. We calculate profiles in
radial shells equally-spaced in $\log\left(r/r_{200}\right)$. As we
discussed in Section~\ref{matching}, haloes of mass above
$10^8~h^{-1}\,M_\odot$ can be bijectively matched in
\textsc{coco-warm} and \textsc{coco-cold}. To compare density profiles
in the two models, we stack the individual profiles of matched and
dynamically relaxed haloes in narrow bins of halo mass of width
$\Delta \log (M_{200}) = 0.3$. To determine whether or not a halo is
relaxed, we make use of the criteria for dynamical equilibrium set
out by \cite{2007MNRAS.381.1450N}: (1) the displacement of the centre
of mass from the potential centre should be less than
$0.07r_{\mathrm{vir}}$ and (2) less than $10\%$ of the mass within
$r_{\mathrm{vir}}$ should be in the form of substructure.

The stacked differential density profiles are shown in
Fig.~\ref{stackden} for a variety of mass bins, with the ratio of the
densities shown in the bottom panels. For masses sufficiently larger
than $\sim~2~\times~10^9~h^{-1}~M_\odot$, we expect negligible
differences in the properties of CDM and WDM haloes: this is apparent
in mass bins with $M_{200} > 10^{11}~h^{-1}\,M_\odot$. Systematic
differences in the density profiles begin to appear at around $M_{200}
\sim 5 \times 10^{10}~h^{-1}\,M_\odot$: the WDM haloes have slightly
but systematically lower central densities than their CDM
counterparts. This halo mass is two orders of magnitude higher than the
half-mode mass, and an order of magnitude higher than the scale at which 
the mass functions begin to differ (Fig.~\ref{sheth}). The difference in central density grows as the
mass decreases and reaches $\sim 30 \%$ at the smallest mass bin
shown, $M_{200} \sim 1.4 \times 10^9~h^{-1}\,M_\odot$. We discuss the
physical reason for this in the next section.

It is now well established that the density profiles of dark matter
haloes in general are well described by the NFW profile
(\citealt{1996ApJ...462..563N,1997ApJ...490..493N}):
\bq
\frac{\rho\left(r\right)}{\rho_c} = \frac{\delta_c}{\left(r/r_s\right)\left(1+r/r_s\right)^2}\;,
\eq
where $\delta_c$ is a characteristic overdensity and $r_s$ is a scale
radius. These two parameters are strongly correlated and depend only
on halo mass (\citealt{1997ApJ...490..493N}). The NFW form is a nearly
universal profile in the sense that it approximately fits the profiles
of relaxed haloes of any mass formed by gravitational instability from all the
initial conditions and cosmological parameters that have been tested
so far. The universality of the NFW profile is intimately related to
the way in which haloes are assembled (\citealt{Ludlow2013}).

We fit NFW profiles to the stacked density profiles of
\textsc{coco-warm} and \textsc{coco-cold} in Fig.~\ref{stackden},
between the radial range defined by the \cite{2003MNRAS.338...14P}
convergence radius, $r_{\mathrm{conv}}$ (defined as the radius within
which the relaxation time is of the order of the age of
the Universe), and $r_{200}$, minimising the following quantity:
\bq
\sigma^2_{\mathrm{fit}} = \frac{1}{N_{\mathrm{bins}}-1} \sum^{N_{\mathrm{bins}}}_{i=1} \left[  \ln \rho_i - \ln \rho_{\mathrm{NFW}} \left( \delta_c;r_{s}\right) \right]^2\;.
\eq
We obtain the best-fitting values of the scale radius, $r_s$, which
defines the halo concentration, $c_{200} = r_{200} / r_{s}$. This
parameter provides a unique characterisation of the NFW density
profile; the values of $c_{200}$ for the stacked profiles are quoted
in Fig.~\ref{stackden}. There is a clear trend in that for large halo
masses, where the density profiles in \textsc{coco-warm} and
\textsc{coco-cold} are similar, the concentrations are nearly
identical but, for masses below $\sim 5 \times
10^{10}~h^{-1}\,M_\odot$, the concentrations of WDM haloes are
systematically lower than those of CDM haloes.

In many cases, even better fits to the density profile are provided by
a formula first used by \cite{1965TrAlm...5...87E} to describe star
counts in the Milky Way. This formula, which has an additional free
parameter, was dubbed the ``Einasto profile'' by
\cite{2004MNRAS.349.1039N}, who showed that it provides a very good
fit to CDM haloes:
\bq
\ln \left( \frac{\rho}{\rho_{-2}}  \right) = -\frac{2}{\alpha} \left[  \left( \frac{r}{r_{-2}}  \right)^\alpha -1   \right]\;,
\eq
where $\rho_{-2}$ is the density at $r = r_{-2}$, the radius at which
the logarithmic slope of the profile is $-2$ (or where $r^2 \rho$ has
its maximum). The parameter $r_{-2}$ in the Einasto profile is
analogous to the scale radius, $r_s$, of the NFW profile. This allows 
an equivalent definition of halo concentration, $c_{200} = r_{200} /
r_{-2}$. The parameter $\alpha$ (not to be confused with the one in Eq.~\ref{alphaMass})
is a shape parameter that controls the curvature of the profile in the inner regions. A value of $\alpha
\simeq 0.17$ results in a good match to CDM haloes over a wide range
of masses \citep{2004MNRAS.349.1039N,2008MNRAS.387..536G}.

This is demonstrated in Fig.~\ref{stackdenEin}, which is similar to
Fig.~\ref{stackden}, but with Einasto profiles fitted instead of NFW
profiles. It is apparent that the shape parameter, $\alpha$, allows a
better fit to the halo density profiles in both \textsc{coco-warm} and
\textsc{coco-cold}, especially in the inner parts. It is also
interesting to note that the concentrations inferred from the Einasto
profile fits tend to be slightly lower than those inferred from the
NFW profile fits especially at higher masses.

\begin{figure}
\includegraphics[scale=0.38]{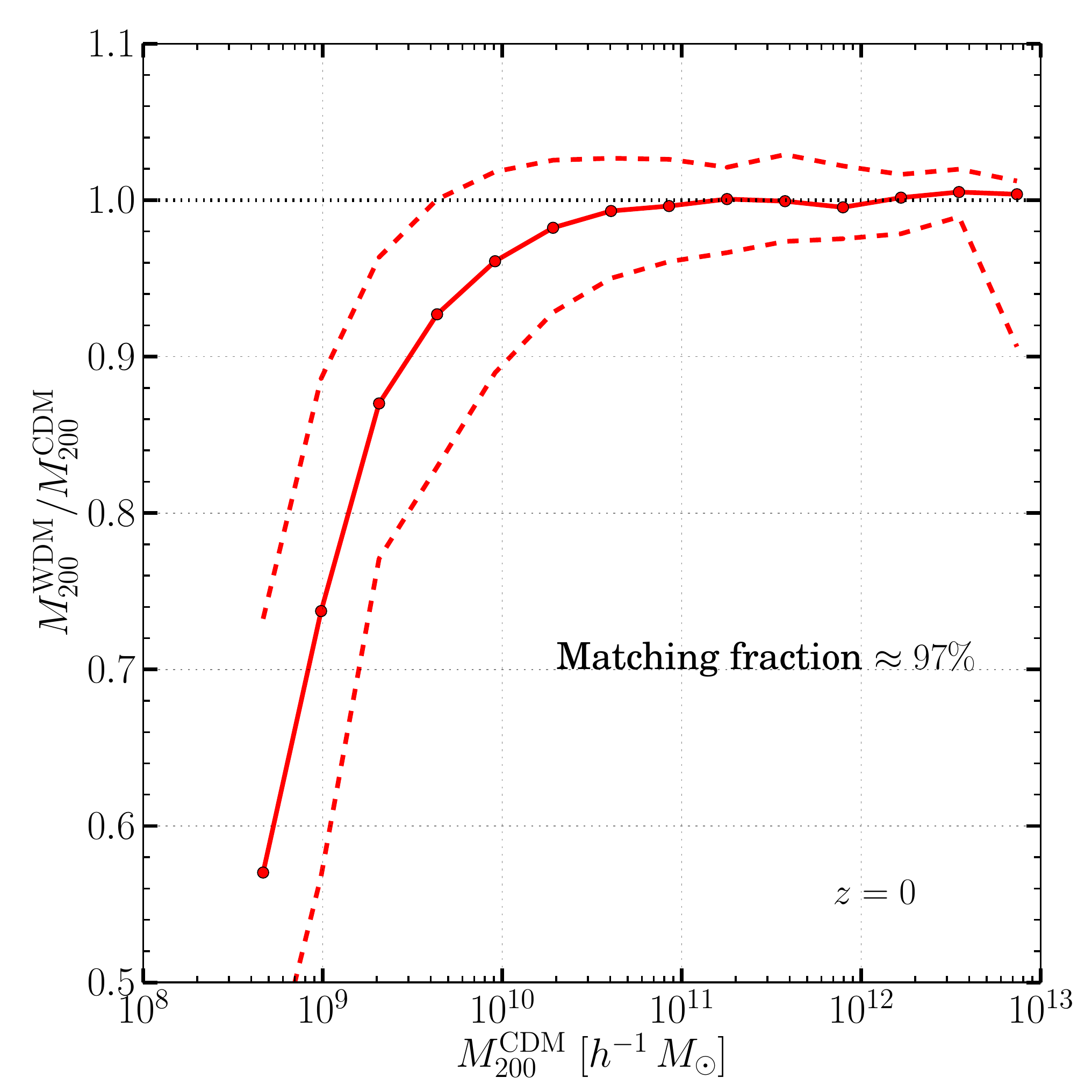}\\
\caption{Ratio of halo mass ($M_{200}$) for all (relaxed and
  unrelaxed) matched haloes above $M_{200} > 10^8~h^{-1}\,M_\odot$ in
  \textsc{coco-warm} and \textsc{coco-cold}, as function of
  $M_{200}^{\mathrm{CDM}}$. The solid red line shows the median relation
  in bins of $M_{200}^{\mathrm{CDM}}$, whereas the dashed red lines indicate the 16-th and 84th percentiles.}
\label{halomassmatch}
\end{figure}

In Fig.~\ref{halomassmatch} we compare the ratio of $M_{200}$ values
for individually matched haloes in \textsc{coco-warm} and
\textsc{coco-cold} at the present day. We consider only haloes with $M_{200} >
10^8~h^{-1}\,M_\odot$ for which we have almost complete matching
($\sim 97 \%$) between the two simulations, and plot the ratio, 
$M_{200}^{\mathrm{WDM}} / M_{200}^{\mathrm{CDM}}$ as a function of
$M_{200}^{\mathrm{CDM}}$. The solid red line shows the median ratio, whereas
the dashed red lines represent the 16-th and 84-th percentiles. The masses are
very similar for objects $> 5 \times 10^{10}~h^{-1}\,M_\odot$, where
the ratios agree to within $1\%$. For masses lower than this, WDM
haloes are systematically less massive than their CDM counterparts,
with the deficit in WDM halo mass reaching $\sim 30 \%$ at
$M_{200}^{\mathrm{CDM}} = 10^{9}~h^{-1}\,M_\odot$. Haloes of
these masses in WDM form later than their CDM counterparts and thus
have less time to grow.

In Fig.~\ref{stacks} we show the cumulative radial distribution of
mass in haloes in \textsc{coco-warm} (red lines) and
\textsc{coco-cold} (blue squares).  The ratios are shown in the lower
panels.  From Fig.~\ref{halomassmatch}, we expect the cumulative
profiles to be very similar at $r/r_{200} = 1$ except in the lowest mass
bin, where WDM haloes are slightly ($\sim 10\%$) less massive than
their CDM matches. The same trend seen in the density profiles is
apparent here: for $M_{200} < 5 \times 10^{10}~h^{-1}\,M_\odot$, the
profiles are less concentrated in the central regions in
\textsc{coco-warm} than in \textsc{coco-cold}. The reason for this
difference is discussed in the next section.

\subsection{The concentration-mass relation}
\label{Conc}

As mentioned in the previous section, the density profile of a dark
matter halo is characterised by its concentration. As a result of
their hierarchical formation process, the inner parts of haloes in CDM
and WDM are essentially in place even before the bulk of the halo mass
is assembled \citep{Wang_2011}. The concentration reflects the mean
density of the Universe at the epoch when these inner regions are in
place and the earlier a halo forms, the higher its concentration is
(\citealt{1997ApJ...490..493N}).

In Section~\ref{density}, we found that the Einasto profile provides a
slightly better fit to the density profiles of WDM and CDM haloes than
does the conventional NFW profile. Furthermore, Einasto fits are less
sensitive to the radial fitting range \citep[][but see also
\citealt{Ludlow2013}]{2008MNRAS.387..536G}.  For these reasons, we
proceed to derive the concentration-mass relation in our simulations
using fits of the Einasto profile to the density profiles of {\it
  individual} haloes (not the stacks). Again, fitting is performed
between the convergence radius, $r_{\mathrm{conv}}$, and $r_{200}$,
while minimising the {\it rms} of the fit:
\bq
\sigma^2_{\mathrm{fit}} = \frac{1}{N_{\mathrm{bins}}-1} \sum^{N_{\mathrm{bins}}}_{i=1} \left[  \ln \rho_i - \ln \rho_{\mathrm{Ein}} \left( \rho_{-2};r_{-2};\alpha \right) \right]^2\;.
\eq

\begin{figure*}
\includegraphics[scale=0.38]{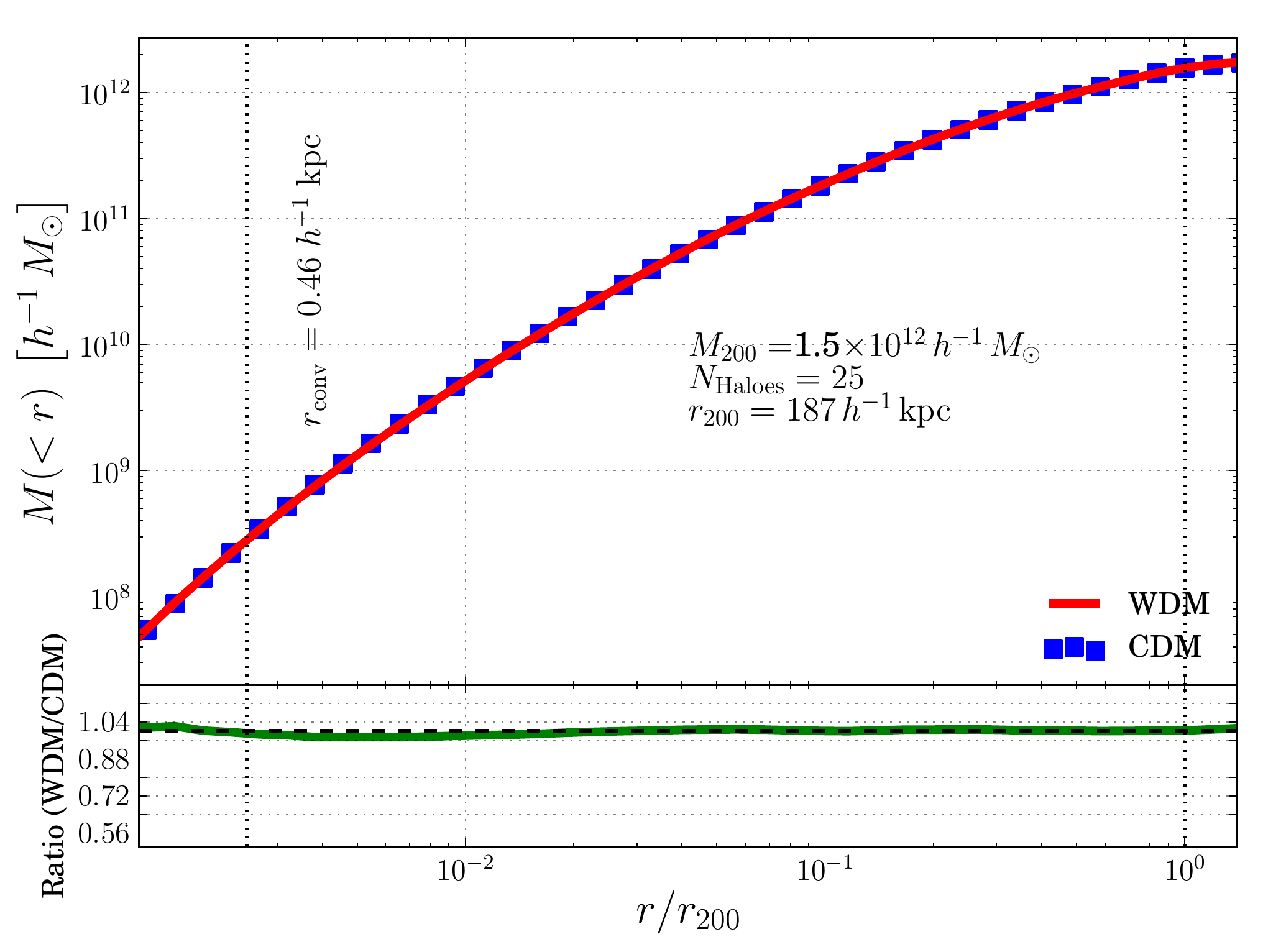}
\includegraphics[scale=0.38]{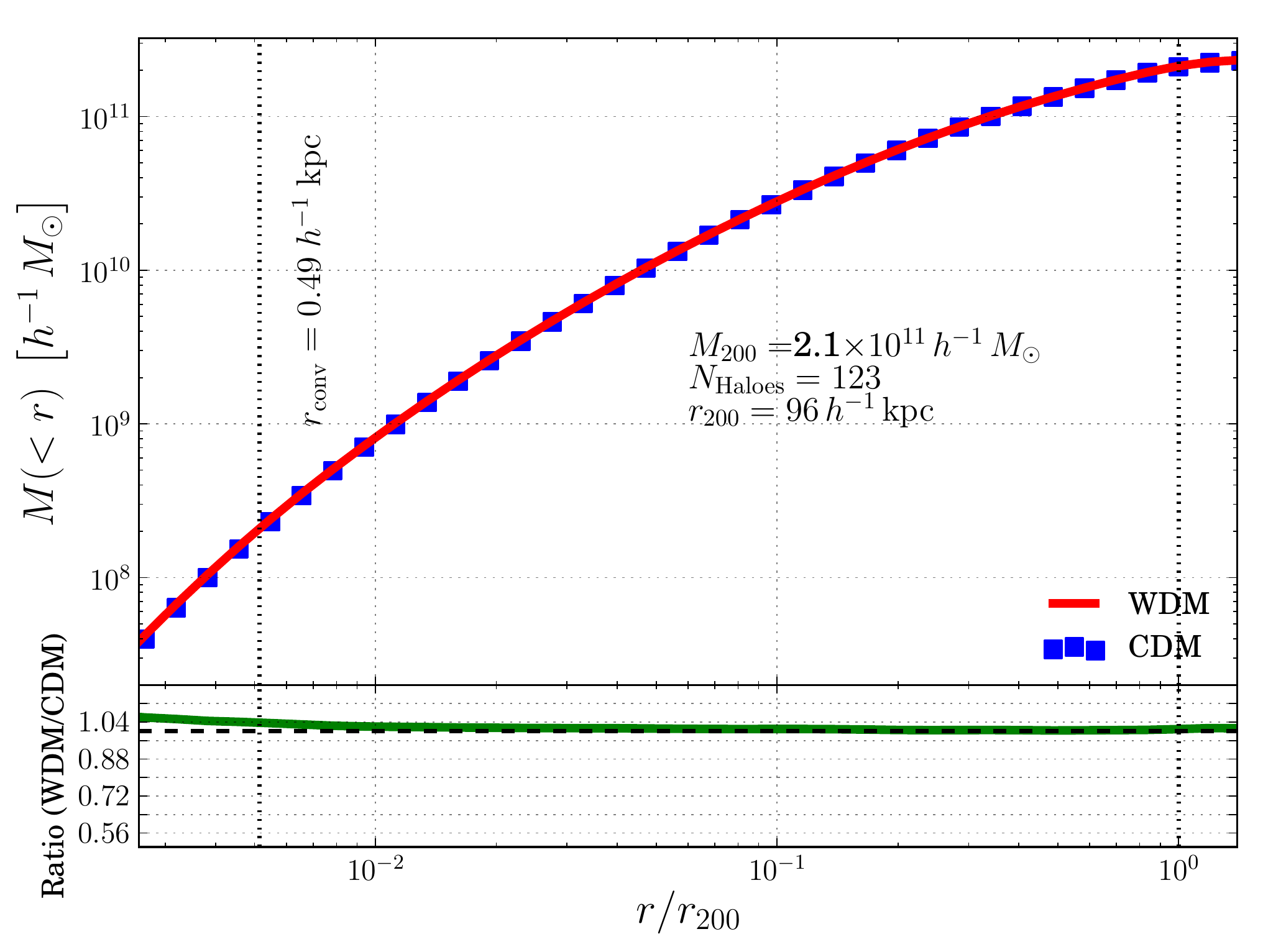} \\ 
\includegraphics[scale=0.38]{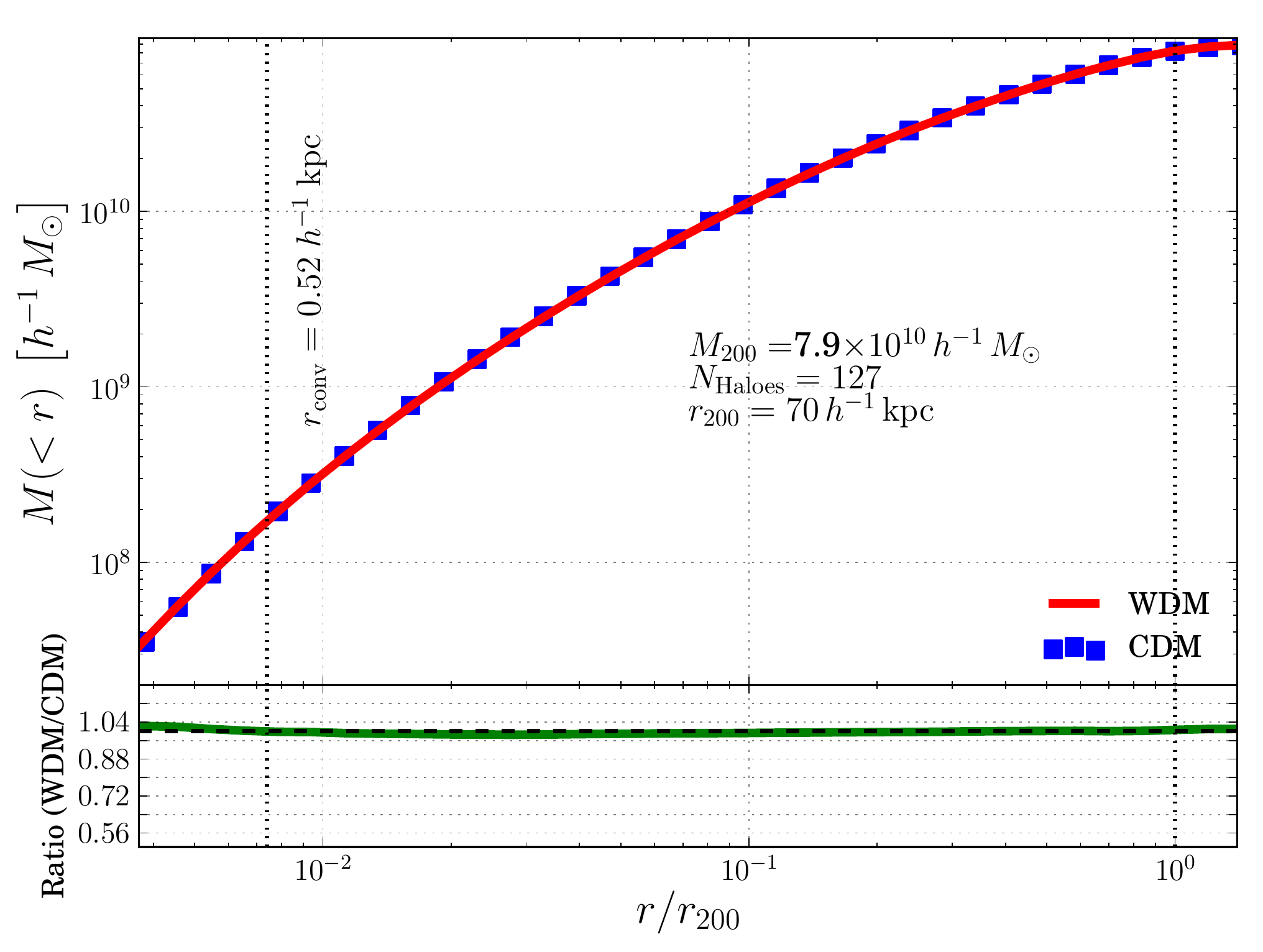}
\includegraphics[scale=0.38]{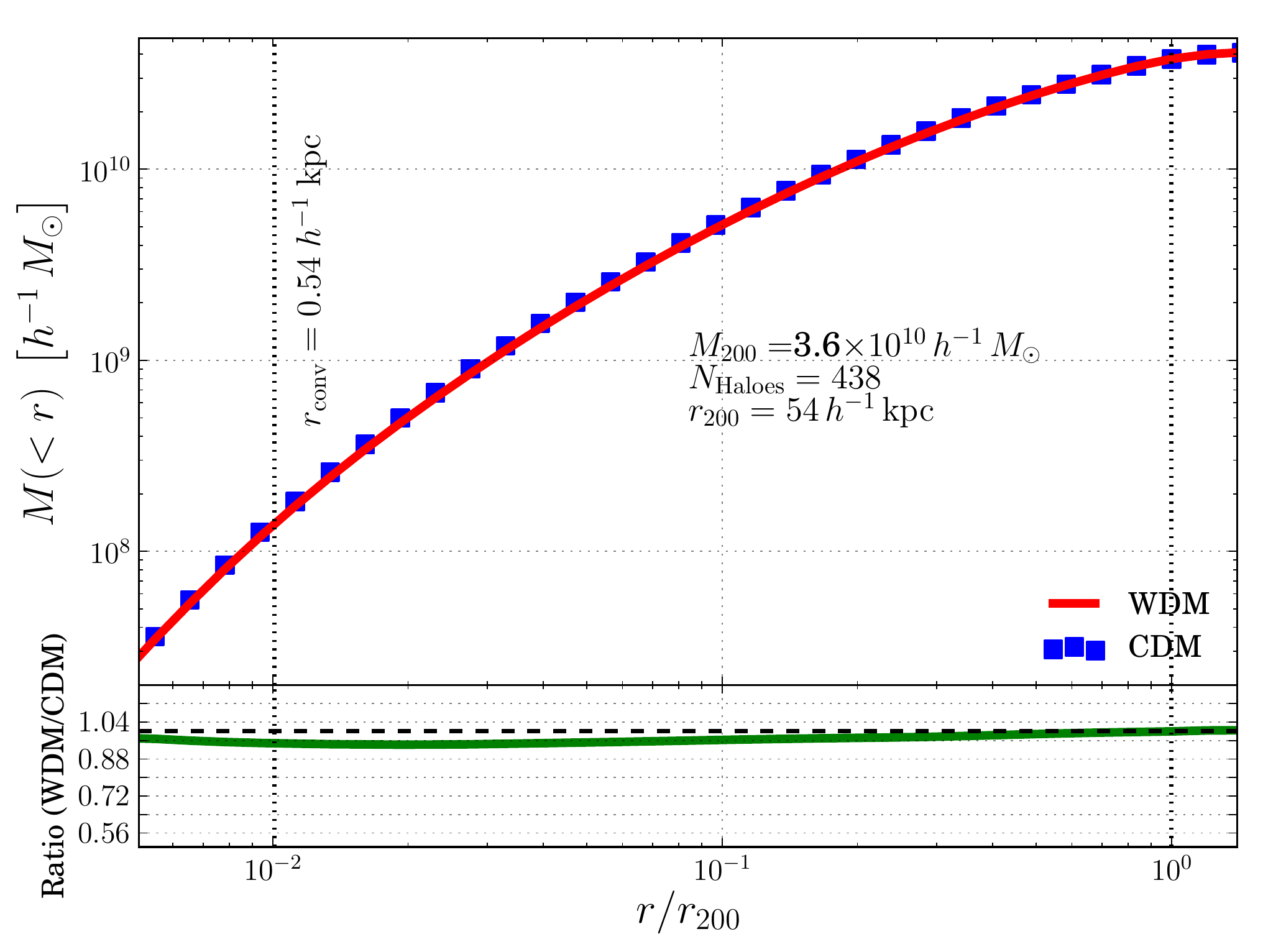}\\
\includegraphics[scale=0.38]{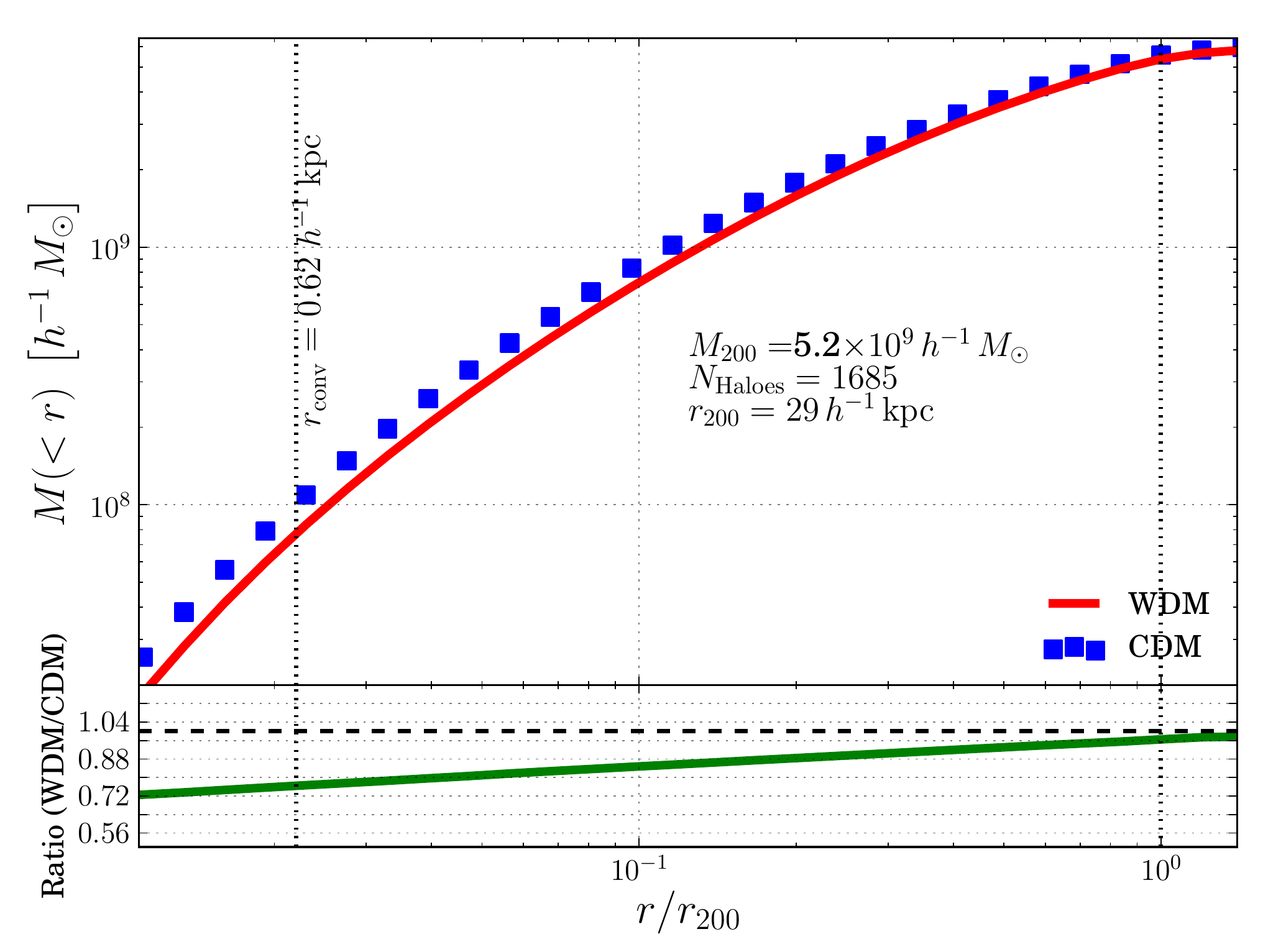}
\includegraphics[scale=0.38]{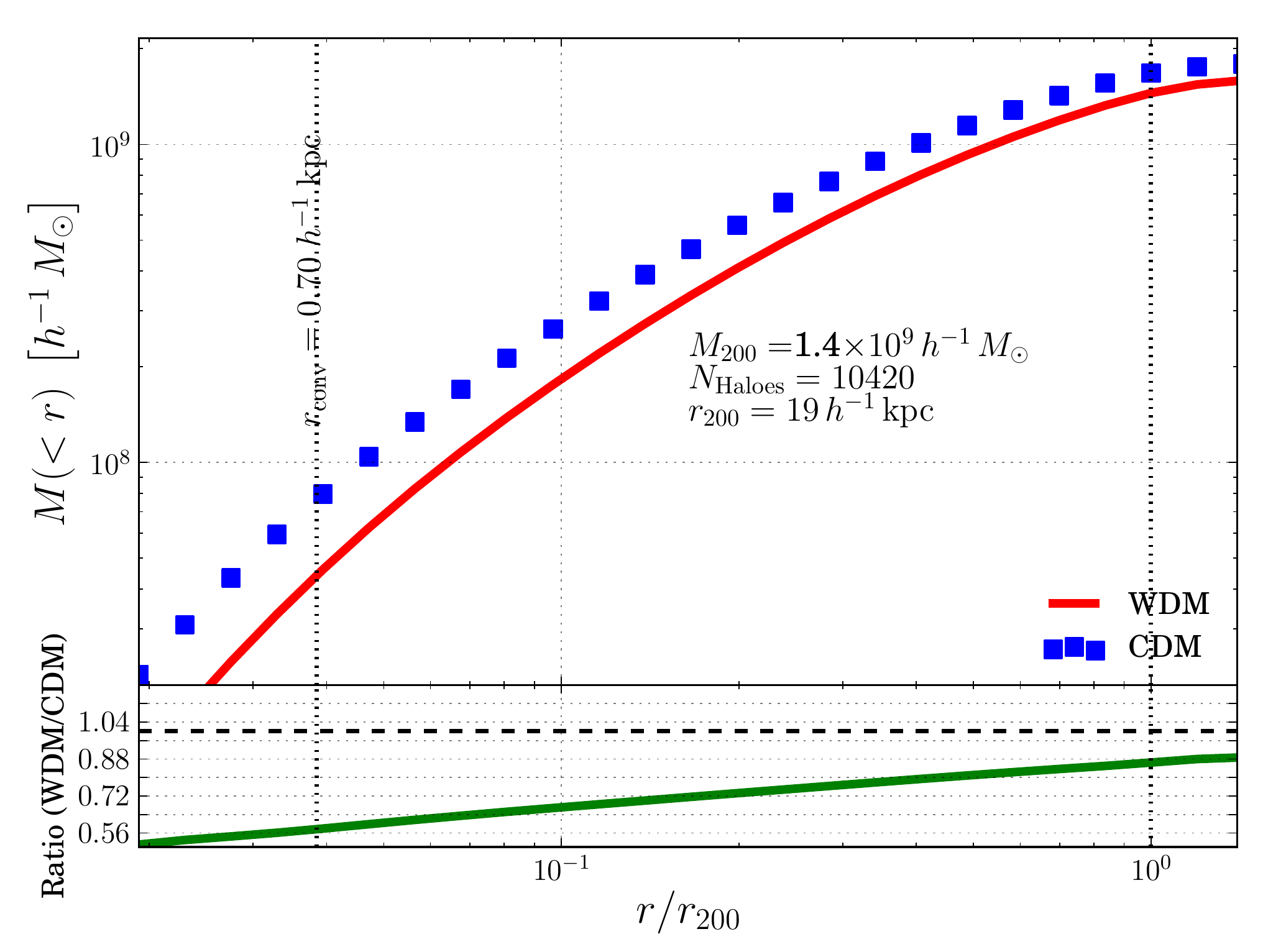}\\
\caption{Stacked cumulative mass profiles of relaxed, matched haloes
  in different mass bins for WDM (solid red lines) and CDM (blue
  squares). The lower panels show the ratio of the WDM mass to the CDM
  mass as a function of radius from the centre of the halo (in units
  of $r_{200}$). For haloes with $M_{200} > 10^{11}~h^{-1}\,M_\odot$,
  the mass profiles are nearly identical, but below $M_{200 }\leq
  5\times 10^{10}~h^{-1}\,M_\odot$ they differ noticeably.}
\label{stacks}
\end{figure*}

To obtain the halo $M_{200}-c_{200}$ relation we first split the
haloes into bins equally spaced in logarithmic mass. We then fit an
Einasto profile to each halo individually, removing all unrelaxed
haloes according to the \cite{2007MNRAS.381.1450N} criteria. We then
find the median halo concentration in each mass bin and estimate its
uncertainty using bootstrap resampling.

\begin{figure}
\includegraphics[width=0.5\textwidth]{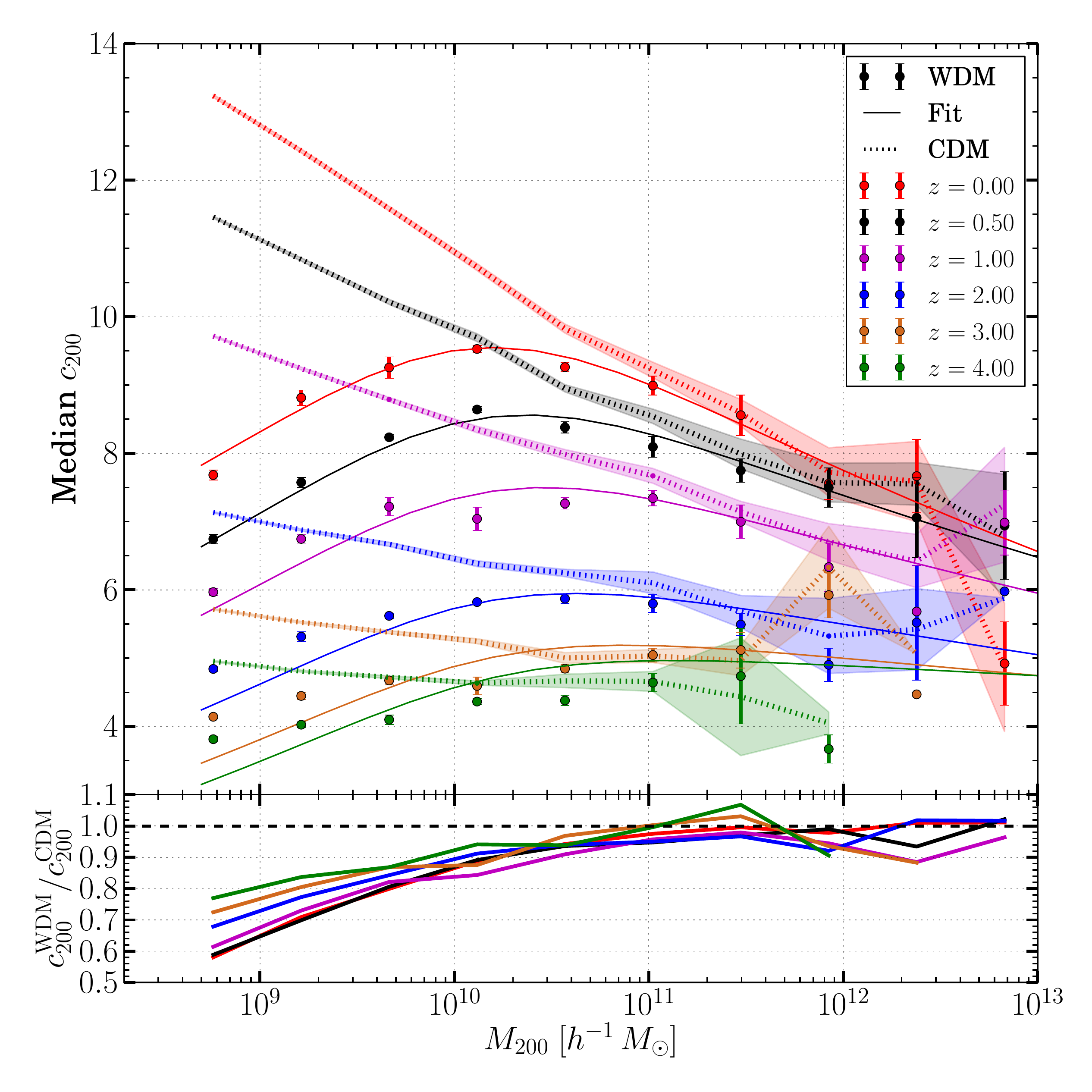}
\caption{The median concentration-mass relation and its redshift
  evolution for haloes in \textsc{coco-cold} and
  \textsc{coco-warm}. The colour dotted lines show the median relation
  over redshift for CDM haloes, as indicated in the legend. The shaded
  regions represent the errors in the median, as estimated by
  bootstrap resampling. The points with the error bars show the
  corresponding redshift relation in WDM. Only relaxed haloes are
  included. The thin colour lines show the results of the fitting formula introduced in Eq.~\ref{fitfunc}.}
\label{conc}
\end{figure}

Fig.~\ref{conc} shows the (median) concentration-mass relations for
\textsc{coco-cold} (dotted lines and shaded regions) and
\textsc{coco-warm} (points with error bars) at redshifts $z=0, 0.5, 1, 2, 3$ and $4$
(different colours as indicated in the legend).  These relations
display the same qualitative behaviour seen in the density profiles in
Fig.~\ref{stacks}. For haloes with mass $M_{200}~>~10^{11}~h^{-1}\,M_\odot$,
the concentrations of CDM and WDM haloes agree well over all redshifts. For masses below this value, WDM haloes
are less concentrated than their CDM counterparts at all
redshifts. This is a direct result of the later formation epoch of
haloes of a given mass in WDM, and reflects the fact that the mass
within $r_{-2}$ in WDM haloes is assembled when the background density
of the Universe is lower than in the CDM case.

Whereas the CDM halo concentrations continue to increase as power laws
towards lower masses, reflecting hierarchical growth, the WDM halo
concentrations turn over at $M_{200}~<~5~\times10^{10}~h^{-1}\,M_\odot$
and eventually begin to decrease \citep[see
also][]{2012MNRAS.424..684S,2013MNRAS.428..882M}.  This echoes the
finding in Fig.~\ref{stacks} that the mass in the central regions of
WDM haloes begins to fall below that in the CDM case roughly below this
mass. This mass is an order of magnitude larger than the mass scale at which the
mass functions begin to differ ($\sim 2 \times 10^9~h^{-1}\,M_\odot$, see
Figs.~\ref{z_form},~\ref{sheth}). This result is not entirely
surprising: the concentration is sensitive to the inner parts of the
profile and it is this inner mass (which we can roughly identify with
the matter contained within $r_{-2}$) which is assembled later in
WDM than in CDM, while most of the mass actually lies in the outer
parts of the halo.  

The lower panel of Fig.~\ref{conc} shows the ratio of the
concentrations in \textsc{coco-warm} and \textsc{coco-cold},
$c_{200}^{\mathrm{WDM}} / c_{200}^{\mathrm{CDM}}$. There are two
interesting features of note: firstly, for all redshifts, the downturn
in the WDM halo concentrations occurs at roughly the same halo mass,
$M_{200} \sim 5 \times 10^{10}~h^{-1}\,M_\odot$; and secondly, at
fixed mass, the ratio decreases with decreasing redshift. The fact that the
mass at which WDM halo concentrations begin to peel-off from the CDM relation is almost
independent of redshift reflects the narrow redshift range in which the inner parts of WDM
haloes form.

In \textsc{coco-warm} we also find that the evolution of the
mass-concentration relation over redshift can be approximated using a
simple functional form motivated by Eq.~\ref{fitzform} 
\citep[see][]{2012MNRAS.424..684S}, with an extra 
redshift-dependent component:
\bq \label{fitfunc}
\frac{c_{200}^{\mathrm{WDM}}}{c_{200}^{\mathrm{CDM}}} = \left(1 + \gamma_1 \frac{M_{\mathrm{hm}}}{M_{200}} \right)^{-\gamma_2} \times \left(1 + z\right)^{\beta(z)}\;.
\eq
Here, $M_{\mathrm{hm}}$ is the half-mode mass, $z$ is the redshift of
interest, $\gamma_1 = 60$, $\gamma_2 = 0.17$ and $\beta(z) = 0.026z -
0.04$. The predictions of our model are shown in the upper panel of
Fig.~\ref{conc} using the thin colour lines. While the model does
not fully capture the relatively flat relationship at $z=3$ and~4 in
\textsc{coco-warm}, it generally reproduces the trends in the
simulation and provides a good fit up to $z=2$, over nearly 5 orders
of magnitude in halo mass.

\subsection{The shapes and spins of haloes}
\label{inertiastuff}

In this section we examine the shapes and spins of WDM haloes. The
shapes are most commonly quantified by the {\it triaxiality}, defined through 
the halo inertia tensor:
\bq \label{MOI}
I_{ij} = m_p \sum^{N_{200}}_{n=1} x_{n,i} x_{n,j}\;,
\eq
where $N_{200}$ is the number of particles within $r_{200}$, $m_p$ is
the mass of the simulation particle, and $x_{n,i}$ is the $i$th
coordinate of the $n-$th particle relative to the halo centre. The
eigenvalues of the inertia tensor define the axial lengths of an
equivalent uniform density ellipsoid, $a \geq b \geq c$, which can be
related to those of the halo itself
(\citealt{2007MNRAS.376..215B}). The sphericity is defined as $c/a$
(as in Section~\ref{spurious}): the higher its value, the less
aspherical the ellipsoid's projection.  The triaxiality is defined as
$T = (a^2 - b^2)/(a^2 - c^2)$: large values correspond to prolate
ellipsoids, small values to oblate ellipsoids. 

\begin{figure}
\includegraphics[width=0.5\textwidth]{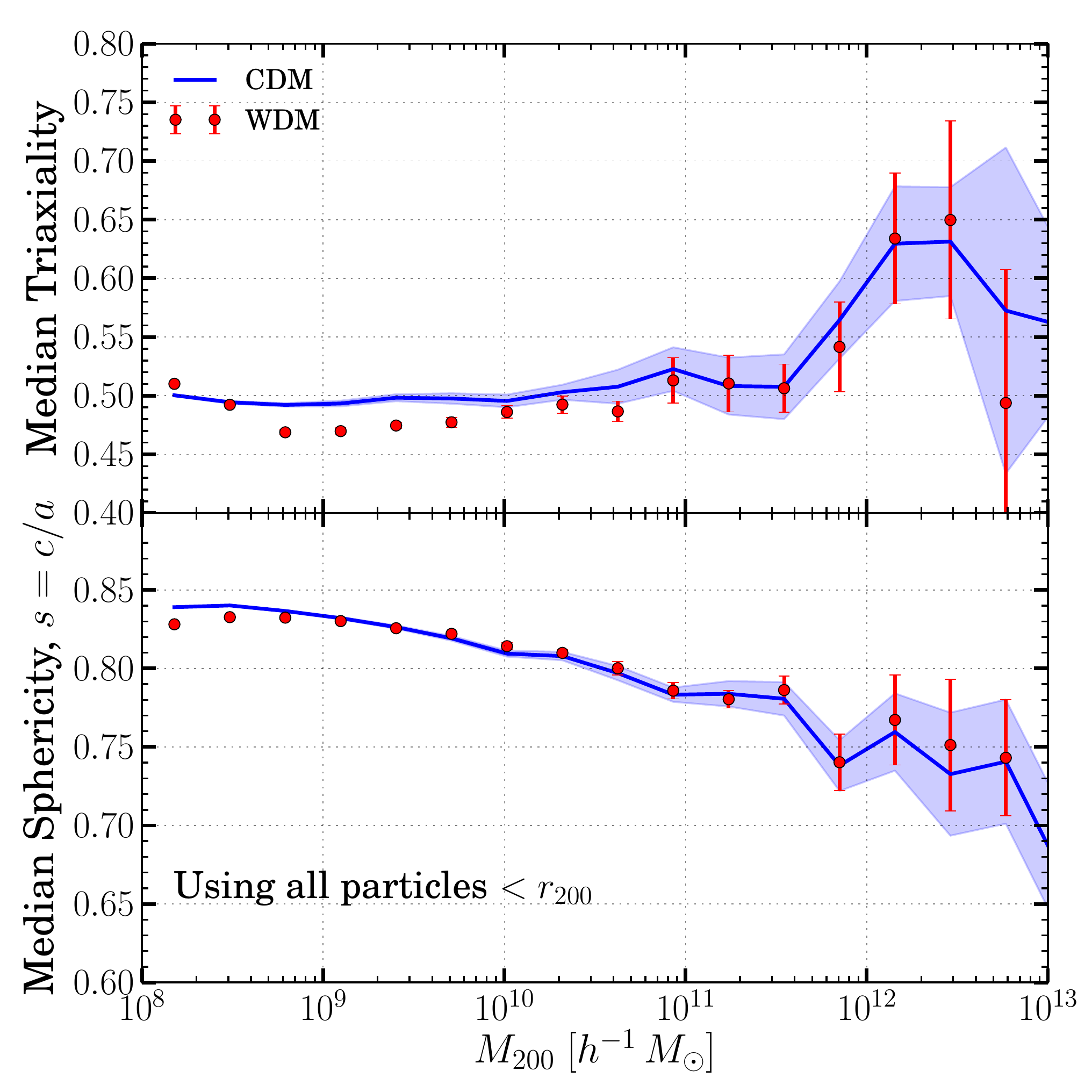}
\caption{Median halo triaxiality (top panel) and halo sphericity
  (lower panel) in \textsc{coco-warm} (red points) and
  \textsc{coco-cold} (blue lines). The errors on the median were
  obtained by bootstrapping 100 different samples in each case, and is
  represented by the red error bars for WDM and the blue shaded region
  for CDM. Only particles within $r_{200}$ were used to compute these
  properties from the inertia tensor.}
\label{sphericitytriax}
\end{figure}

The results for our simulations are shown in
Fig.~\ref{sphericitytriax}, where blue represents CDM and red WDM,
with the top panel comparing the median triaxiality, and the lower the
median sphericity. Errors on the median quantities were obtained by
bootstrap resampling. Previous $N$-body simulations of CDM haloes have
shown that triaxiality correlates with halo mass, with triaxility
decreasing with decreasing halo mass
(\citealt{Frenk88,2006MNRAS.367.1781A,2011MNRAS.411..584M,2013MNRAS.428..882M}).
This trend reflects, in part, the younger dynamical age of more
massive haloes. Fig.~\ref{sphericitytriax}
shows that the same trend is present for WDM haloes but below $M_{200}
\sim 10^{10}~h^{-1}\,M_\odot$, WDM haloes are slightly less triaxial
than their CDM counterparts.

A more significant trend is revealed when comparing the spin of haloes
in the two simulations. The spin is best characterised by the
parameter, $\lambda$, defined as:
\bq\label{spineq}
\lambda = \frac{J\sqrt{\left|E\right|}}{GM^{5/2}}\;
\eq 
\citep{1969ApJ...155..393P}, where $J$ is the magnitude of the angular
momentum of the halo, $E$ is its total energy and $M$ is the mass
(which we take to be $M_{200}$). Haloes acquire a net angular momentum
through tidal torques during growth in the linear regime which can be
subsequently modified and rearranged by mergers
\citep{1969ApJ...155..393P,1970Ap......6..320D,1984ApJ...286...38W}. Since the
merger histories are different for CDM and WDM haloes, we might expect
some differences in their final angular momentum configurations.  In
particular, given that tidal forces associated with mergers tend to
redistribute angular momentum from the central parts of haloes to the
rest of the halo, the smaller frequency of mergers in WDM might
facilitate the formation of extended spinning galactic disks
\citep{Frenk88,1991ApJ...380..320N,1994MNRAS.267..401N}
\footnote{We note that the inability of many early simulations to form
  extended disks in the CDM model -- the so-called ``angular
  momentum'' problem -- is readily solved when appropriate
  prescriptions for supernovae feedback are included in the
  simulations \citep[see e.g.][]{2005MNRAS.363.1299O,
    Scannapieco_2011}. }.

\begin{figure}
\includegraphics[width=0.5\textwidth]{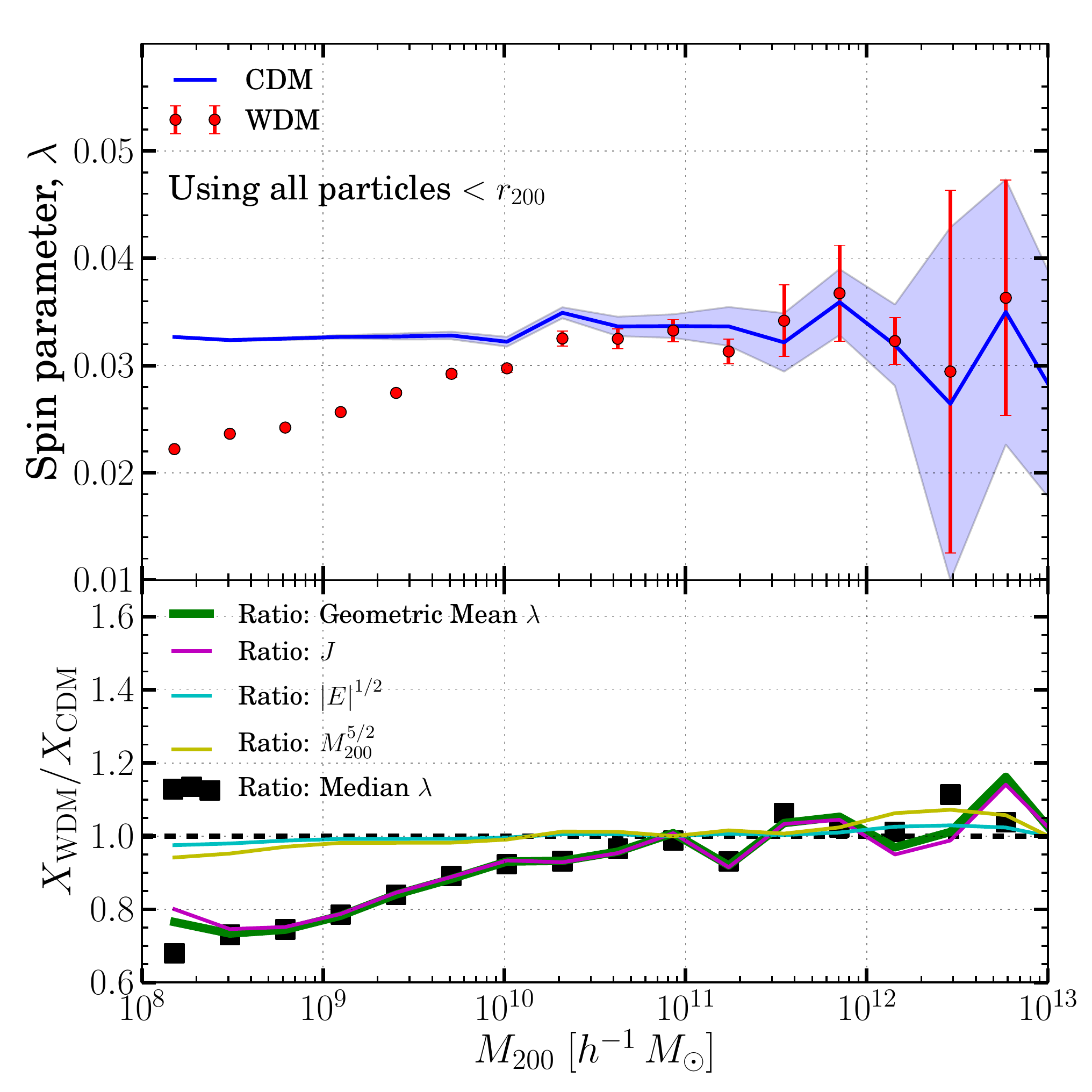}
\caption{Top panel: the median halo spin-mass relation at $z=0$ for
  \textsc{coco-warm} (red points) and \textsc{coco-cold} (blue
  line). Errors on the median for the WDM (shown by error bars) and
  for CDM (shown as the shaded region) haloes were calculated by
  bootstrap resampling. Bottom panel: the relative contributions of
  energy, angular momentum and halo mass to the spin of the halo. The
  black squares show the ratio (CDM to WDM) of the median spin
  parameters (from the top panel). The magenta, cyan and yellow lines
  show the ratios of the geometric means of the angular momentum,
  energy and $M_{200}$ respectively, which when multiplied together
  appropriately yield the thick green line, which show the ratio of the
  geometric means of $\lambda_{\mathrm{CDM}}$ and
  $\lambda_{\mathrm{WDM}}$. As expected, the squares trace out the
  ratio of the geometric means. Note that ratios of all quantities are taken between the bijectively matched 
  \textsc{coco-warm} and \textsc{coco-cold} haloes.}
\label{spins}
\end{figure} 

The spin parameters in our two simulations are compared in the top
panel of Fig.~\ref{spins}.  Previous cosmological CDM simulations
showed a very weak correlation between spin and halo mass, with a
median value of $\lambda \approx 0.033$, across a wide range of halo
masses \citep{1985ApJ...292..371D,Barnes87,spin1,spin4,1996MNRAS.281..716C,spin2,spin3,2007MNRAS.376..215B}.
Our \textsc{coco-cold} simulation reproduces this trend and extends it
to lower masses, $M_{200}=10^8~h^{-1}\,M_\odot$.  

For $M_{200} > 5 \times 10^{10}~h^{-1}\,M_\odot$, the $\lambda$ values
for WDM haloes are almost identical to those of their CDM
counterparts.  However, for smaller halo masses $\lambda$ decreases
systematically with decreasing mass and is lower than the CDM value by
almost 30\% at $M_{200} \sim 10^8~h^{-1}\,M_\odot$. This is consistent
with the results of \cite{2002ApJ...564L...1B}, who found that three
out of four haloes below the WDM cutoff in their simulation had lower
values of $\lambda$ than the equivalent CDM matches.  Note that in the top panel of
Fig.~\ref{spins} we include all haloes, not necessarily matches,
which explains why in some of the largest mass bins, the median spins
are not exactly the same in WDM and CDM.  In addition, we only include
haloes with more than 1000 particles within $r_{200}$ since particle
shot noise dominates the estimates of angular momentum for low
particle numbers (\citealt{Frenk88,2007MNRAS.376..215B}, although
we use a more conservative lower limit than the latter's choice of 300
particles).

To investigate why the spins of dwarf galaxy haloes are lower in WDM
than in CDM we consider the relative contributions of energy, angular
momentum and $M_{200}$ to $\lambda$, illustrated in the bottom panel
of Fig.~\ref{spins}, this time for bijectively matched haloes. The ratio of the median spin parameters is
shown by the black squares and the ratio of the geometric means of
the quantities that enter into Eq.~\ref{spineq} are shown by the other
colour lines (magenta for $J_{\mathrm{CDM}} / J_{\mathrm{WDM}}$, cyan
for $\left|E_{\mathrm{CDM}} / E_{\mathrm{WDM}}\right|^{1/2}$ and yellow for $\left(M_{200,
  \mathrm{CDM}} / M_{200, \mathrm{WDM}}\right)^{5/2}$). The combination of these
ratios in Eq.~\ref{spineq} should reproduce the ratio of spin
parameters, and this is shown in the thick green line.  Part of the
reason for lower WDM spins below $\sim 10^{10}~h^{-1}\,M_\odot$ is
their slightly lower total energy which results from their lower
concentration. The dominant factor, however, is their lower angular
momentum relative to CDM haloes, $\sim 25\%$ at
$10^{8}~h^{-1}\,M_\odot$.  The cause of this could be related to the
differing merger histories in WDM and CDM and the likely more
quiescent mass accretion of WDM haloes which can result in smaller
spins \citep{2002ApJ...564L...1B,2002ApJ...581..799V,2006MNRAS.370.1905H}.

\section{Conclusions}
\label{conclusion}

We have presented results from the {\it COpernicus COmplexio} project,
a set of cosmological ``zoom'' simulations in which the dark matter is
assumed to be either CDM (\textsc{coco-cold}) or a thermal 3.3 keV WDM
particle (\textsc{coco-warm}). The combination of mass resolution and
volume of our simulations provides a rich statistical sample of haloes
over seven decades in mass. This WDM model is particularly interesting
because it corresponds to the ``warmest'' particle allowed by current
Lyman$-\alpha$ constraints \citep{2013PhRvD..88d3502V} and has a linear power
spectrum cutoff similar to that for the ``coldest'' 7 keV sterile
neutrino, evidence for which has recently been claimed to be found in galaxies and
clusters \citep{2014ApJ...789...13B, 2014PhRvL.113y1301B}. This cutoff -- manifest in haloes of $M_{200} \leq 2\times
10^{9}~h^{-1}\,M_\odot$ for our assumed particle mass -- is reflected
both in the population statistics and the structure of individual
haloes.

The formation of structure begins significantly later in
\textsc{coco-warm} than in \textsc{coco-cold}. Across all redshifts,
differences in the halo mass function between \textsc{coco-warm} and
\textsc{coco-cold} begin to appear at a mass roughly one order
magnitude larger than the nominal half-mode mass. Below
$\sim 2 \times 10^{9}~h^{-1}\,M_\odot$, the WDM mass function declines rapidly
but there are still some small haloes present at suprisingly large
redshifts: at $z=10$, for example, there are almost 5 times as many
haloes with $M_{200} \sim 10^8~h^{-1}\,M_\odot$ in \textsc{coco-cold} than in
\textsc{coco-warm}.  We find that the $z =0$ halo mass functions in
both \textsc{coco-warm} and \textsc{coco-cold} are well described by
previous analytic fits to the CDM halo mass function
\citep[e.g.][]{1999MNRAS.308..119S} down to our resolution limit,
$M_{200} \sim 3 \times 10^7~h^{-1}\,M_\odot$, provided that the window function
used to compute the mass variance, $\sigma^2(M)$, in the WDM case is
calculated using a sharp $k$-space filter, as described by
\cite{2013MNRAS.428.1774B}.

Just as for \textsc{coco-cold}, the spherically averaged density
profiles of haloes in \textsc{coco-warm}, down to dwarf galaxy scales,
are well described by NFW or Einasto profiles. The concentration-mass
relation, $M_{200} - c_{200}$ (where we have defined concentration
using the Einasto profile), in \textsc{coco-warm} begins to peel off
from the corresponding relation in \textsc{coco-cold} at a mass of
$\sim 5\times 10^{10}~h^{-1}\,M_\odot$, reflecting the later formation
epoch of haloes of a given mass in WDM compared to CDM. This mass is
larger than the scale below which the WDM mass function is suppressed because halo
concentration is determined by the epoch when the {\em inner} regions
of a halo form. The mass at which the concentration begins to differ
in the two simulations is almost independent of redshift out to
$z\simeq 4$. At the present day, the typical concentration of a halo
of mass $10^9~h^{-1}\,M_\odot$ in \textsc{coco-warm} is $c_{200}\simeq
8$ compared to $c_{200}\simeq 12.7$ in \textsc{coco-cold}. The trends
and evolution of the concentration-mass relation can be approximated
by the fitting formula provided in Eq.~\ref{fitfunc}.

The generally triaxial shapes of haloes in \textsc{coco-warm} and
\textsc{coco-cold} are very similar. However, we find that, for masses
below $\sim 5\times 10^{10}~h^{-1}\,M_\odot$, WDM haloes have slightly
lower values of the spin parameter, $\lambda$, (up to 30\%) than their
CDM counterparts. Whether this and the other differences in halo
properties between WDM and CDM that we have found have implications
for galaxy formation needs to be investigated using hydrodynamic
simulations.

\section*{Acknowledgements}

We would like to thank Maciej Cytowski and Arkadiusz Niegowski for
overseeing the \textsc{coco-cold} simulations at University of Warsaw
HPC centre, and to Oleg Ruchayskiy and Alexey Boyarsky for providing us with the sterile neutrino power spectra used in Fig.~\ref{matterps}. We also greatly appreciate many useful comments and
feedback from Till Sawala over the course of this project. SB is
supported by STFC through grant ST/K501979/1. This work was supported
in part by ERC Advanced Investigator grant COSMIWAY [grant number GA
267291] and the Science and Technology Facilities Council [grant
number ST/F001166/1, ST/I00162X/1]. WAH is also supported by the
Polish National Science Center [grant number
DEC-2011/01/D/ST9/01960]. BL is supported by the Royal Astronomical
Society and Durham University. This work used the DiRAC Data Centric
system at Durham University, operated by the Institute for
Computational Cosmology on behalf of the STFC DiRAC HPC Facility
(\url{www.dirac.ac.uk}). This equipment was funded by BIS National
E-infrastructure capital grant ST/K00042X/1, STFC capital grant
ST/H008519/1, and STFC DiRAC Operations grant ST/K003267/1 and Durham
University. DiRAC is part of the National E-Infrastructure. This
research was carried out with the support of the HPC Infrastructure
for Grand Challenges of Science and Engineering Project, co-financed
by the European Regional Development Fund under the Innovative Economy
Operational Programme. This work is also part of the D-ITP consortium,
a program of the Netherlands Organisation for Scientific Research
(NWO) that is funded by the Dutch Ministry of Education, Culture and
Science (OCW). The data analysed in this paper can be made available upon request 
to the author.


\bibliographystyle{mnras}
\bibliography{bibl_WAH2.bib}{}

\label{lastpage}

\end{document}